\documentclass[aps,prd,twocolumn,superscriptaddress,preprintnumbers,nofootinbib,showpacs]{revtex4-1}

\usepackage{graphicx}
\usepackage{dcolumn}
\usepackage{bm} 

\usepackage[english]{babel}

\usepackage{hyperref}

\usepackage[utf8]{inputenc}

\usepackage{color}

\newcommand{\be}{\begin{equation}}
\newcommand{\ee}{\end{equation}}
\newcommand{\bey}{\begin{eqnarray}}
\newcommand{\eey}{\end{eqnarray}}
\newcommand{\tr}{\text{Tr}}

\begin{document} 
\title{Renormalization of the Polyakov loop with gradient flow}
\author{P.~Petreczky}
\affiliation{Physics Department, Brookhaven National Laboratory, Upton, New York 11973, USA}
\author{H.-P.~Schadler}
\affiliation{Physics Department, Brookhaven National Laboratory, Upton, New York 11973, USA}
\affiliation{Institute of Physics, University of Graz, 8010 Graz, Austria}

\date{\today}

\begin{abstract}
We use the gradient flow for the renormalization of the Polyakov loop in various representations. 
Using 2+1 flavor QCD with highly improved staggered quarks and lattices with temporal extents of $N_\tau=6$, $8$, $10$ and $12$ we 
calculate the renormalized Polyakov loop in many representations including fundamental, sextet, adjoint, decuplet, 
15-plet, 24-plet and 27-plet. This approach allows for the calculations of the renormalized Polyakov loops over a large temperature range 
from $T=116$~MeV up to $T=815$~MeV, with small errors not only for the Polyakov loop in fundamental representation, 
but also for the Polyakov loops in higher representations. 
We compare our results with standard renormalization schemes and discuss the Casimir 
scaling of the Polyakov loops.
\end{abstract}


\maketitle
 
\section{Introductory remarks}

The Polyakov loop in fundamental representation is an order parameter
of deconfinement in SU(N) gauge theories.
For the SU(3) gauge group it is defined as
\begin{equation}
	L_3(\mathbf{x}) = \frac{1}{3} \mbox{Tr} \, {\cal P}  \exp \bigg( \int_0^{1/T} \! A_4^a(\mathbf{x},\tau) t^a\, d\tau \bigg) \; ,
\label{def}
\end{equation}
where $\mathbf{x}$ is the spatial coordinate, ${\cal P}$ is the path ordering operator, and the Euclidean time $\tau$ is integrated up to the inverse temperature. 
The nonzero expectation value of $L_3(\mathbf{x})$
above the transition temperature $T_c$ signals deconfinement and screening of static quarks.
After proper renormalization the logarithm of the Polyakov loop gives the free 
energy of a static quark in temperature units \cite{McLerran:1981pb,Kaczmarek:2002mc}.
In the confining phase below $T_c$ the corresponding free energy is infinite. 
Above that temperature it becomes finite due to color screening.

In QCD the Polyakov loop is not an order parameter, its expectation value is nonzero at any temperature as static quarks can be screened
by dynamical quarks already in the vacuum, i.e. the free energy of the static quark is always finite. Nonetheless, the temperature dependence of
the Polyakov loop reflects the change of the screening properties in the medium and thus is linked to deconfinement.

So far, we discussed the Polyakov loop in the fundamental representation. One can define the Polyakov loop $L_n(\mathbf{x})$ in any representation $n$ by replacing
the generators $t^a$ of the fundamental representation by the generators of the corresponding representation $t_n^a$, 
as well as the corresponding normalization of the trace in Eq.~(\ref{def}),
and consider the free energy of the color charge in representation $n$. The color charges in higher representations may be
screened at any temperature already in pure gauge theory. However, also in this case the temperature dependence of 
$\langle L_n(\mathbf{x}) \rangle$ ,
or  equivalently of the corresponding free energy $F_n$, is sensitive to the screening properties of the medium and thus to 
deconfinement.

As stated above, the expectation value of the Polyakov 
loop $P_n(T)=\langle L_n(\mathbf{x}) \rangle$ requires renormalization in order to be interpreted as the free
energy of static charges. The renormalization of the Polyakov loop is multiplicative \cite{Polyakov:1980ca}
\begin{equation}
        P_n(T) \equiv P^{\rm{ren}}_n(T) = e^{-e_n(g_0)N_\tau} P^{\rm{bare}}_n(T) =Z_n^{N_\tau} P^{\rm{bare}}_n(T) \; ,
\label{renorm}
\end{equation}
where $g_0$ is the bare gauge coupling corresponding to a given lattice spacing\cite{Kaczmarek:2002mc}.
In the fundamental representation the renormalization is usually achieved by requiring that the free
energy of a static quark antiquark pair is equal to the corresponding zero temperature potential
at very short distances and assuming a certain normalization of the zero temperature potential.
The constant $e_3(g_0)$ in Eq.~(\ref{renorm}) corresponds to the additive shift of the 
zero temperature potential ensuring that it has the prescribed value in physical units.
Thus the calculation of the renormalized Polyakov loop requires the calculation of the zero
temperature potential for each value of the bare gauge coupling $g_0$ used in finite temperature
calculations. For higher representations one can proceed in a similar
manner to obtain the renormalization constant $Z_n$ but usually, as we discuss later,
the assumption of Casimir scaling is used to estimate them.

The renormalized Polyakov loop in the fundamental representation has been calculated
in SU(N) gauge theories \cite{Kaczmarek:2002mc,Digal:2003jc,Gupta:2007ax,Mykkanen:2012ri} as well
as in QCD with two and three quark flavors with relatively large quark masses \cite{Kaczmarek:2005ui,Petreczky:2004pz}.
Results for the renormalized Polyakov loop also exist for the physically relevant case of 2+1 flavor
QCD with physical or nearly physical quark 
masses \cite{Cheng:2007jq,Bazavov:2009zn,Aoki:2006br,Aoki:2009sc,Borsanyi:2010bp,Bazavov:2011nk,Bazavov:2013yv}.
The Polyakov loop in higher representations has also been studied in pure gauge 
theory \cite{Gupta:2007ax,Mykkanen:2012ri} and in two-flavor QCD 
with relatively large quark masses \cite{Gupta:2007ax}. 

In this paper we calculate the renormalized Polyakov loop in 2+1 flavor QCD with physical quark masses 
in various representations. Calculations will be performed at several lattice spacings in order to
control the discretization effects. A new method for calculating the renormalized Polyakov loops
based on the gradient flow \cite{Luscher:2010iy} is introduced. 
The rest of the paper is organized as follows: In the next section we discuss
the lattice setup. In Sec. III we discuss the renormalization of the Polyakov loop in the fundamental representation using
the gradient flow. The Polyakov loop in higher representations and Casimir scaling is studied in Sec. IV.  Finally, Sec. V contains
our conclusions and some technical aspects of the calculations are presented in the Appendices.
\begin{figure*}
	\includegraphics[width=0.45\textwidth,clip]{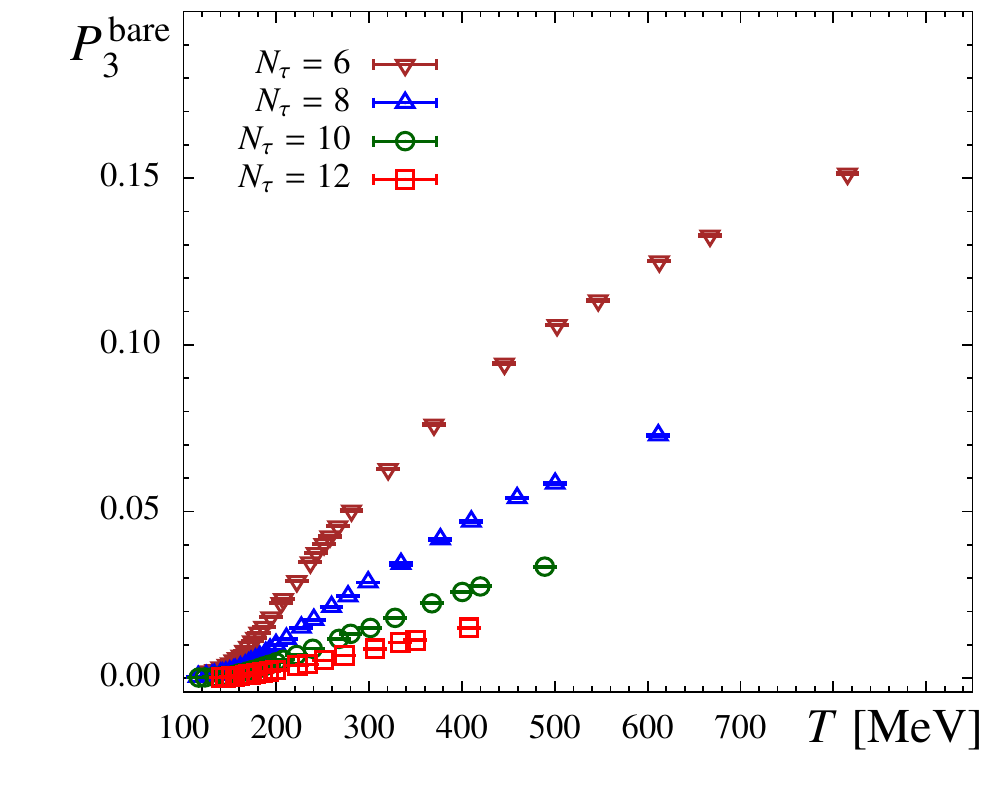}
  \includegraphics[width=0.45\textwidth,clip]{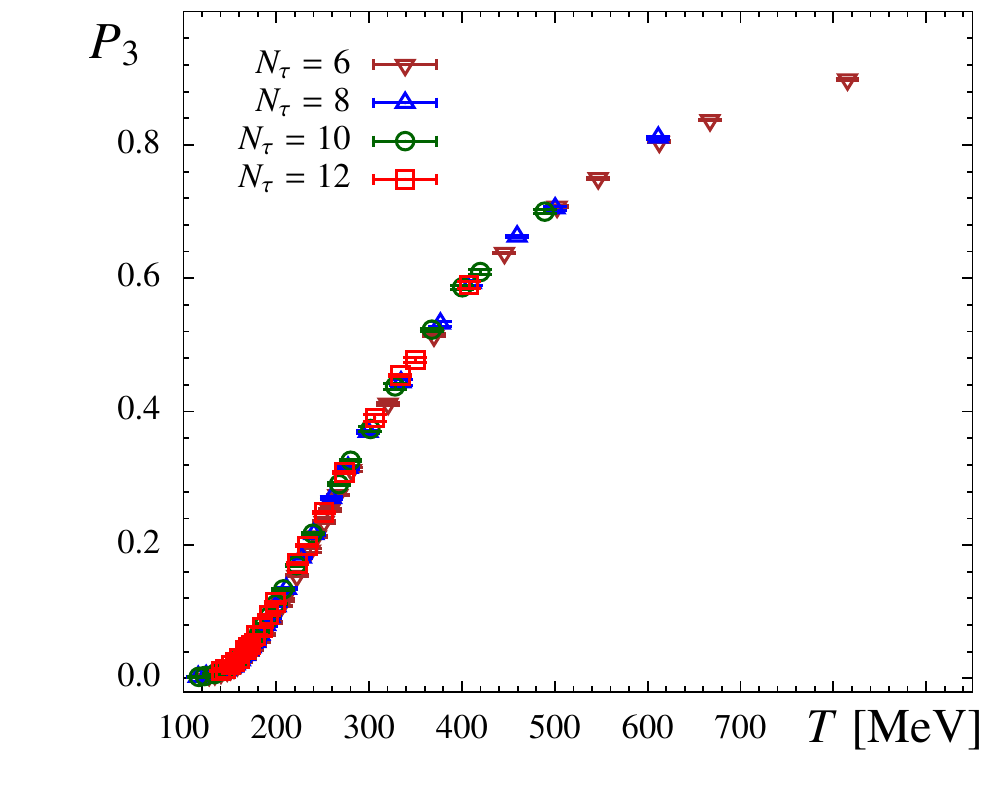}
	\caption{The unrenormalized fundamental Polyakov loop (left) and
         the renormalized Polyakov loop corresponding to flow time $f=f_0$ (right)
          as a function of the temperature $T$ for all lattice ensembles.}
	\label{fig1}
\end{figure*}

\section{Lattice setup}

We perform lattice calculations in 2+1 flavor QCD using highly improved staggered quarks (HISQ) \cite{Follana:2006rc} with lattice 
sizes of $N_{s}^3 \times N_{\tau}$=$24^3 \times 6$, $32^3 \times 8$, $40^3 \times 10$ and $48^3 \times 12$ 
and gauge configurations generated by the HotQCD collaboration using a physical strange quark mass $m_s$ and degenerate 
up and down quarks with masses $m_u=m_d \equiv m_l=m_s/20$ \cite{Bazavov:2011nk,Bazavov:2014pvz}. 
This setup corresponds to a pion mass of $161$~MeV in the continuum limit \cite{Bazavov:2011nk}. 
The temperature $T=1/(a N_\tau)$ is varied using the lattice spacing $a$ and the lattice spacing itself 
has been determined using the $r_1$ scale defined in terms of the static quark potential
\begin{equation}
	\left.r^2 \frac{dV}{dr}\right|_{r=r_1}=1 \; .
\end{equation}
We use the parametrization of $a/r_1$ from Ref.~\cite{Bazavov:2014pvz} and $r_1=0.3106$~fm 
to convert to physical units. We will cover a temperature range of $T=116$~MeV up to $T=815$~MeV.
In the low temperature region and in the transition region we also perform 
calculation using $m_l=m_s/40$ and the HotQCD gauge configurations from Ref. \cite{Bazavov:2011nk}.

On the lattice the local Polyakov loop $L(\mathbf{x})$ is given by the traced product 
of all temporal links $U_4(\mathbf{x},\tau)$ at the spatial point $\mathbf{x}$
\begin{equation}\label{eq:pl}
	L^{\rm{bare}}_3({\mathbf{x}}) = \frac{1}{3} \tr \prod_{\tau=1}^{N_\tau} U_4(\mathbf{x},\tau) \; ,
\end{equation}
with $U_4(\mathbf{x},\tau) \in$~SU(3) and $N_\tau$ the temporal extent of the lattice. 
Here $L^{\rm{bare}}_3({\mathbf{x}})$ denotes the unrenormalized (bare) Polyakov loop  in fundamental (``3'') representation. 
One usually considers the spatial average when calculating the expectation value of the Polyakov loop,
\begin{equation}\label{eq:plspatial}
	P^{\rm{bare}}_3 \; = \left\langle  \frac{1}{N_s^3} \sum_\mathbf{x} L^{\rm{bare}}_3(\mathbf{x})  \right\rangle \; .
\end{equation}
We consider  Polyakov loops $L_n$ in higher representations, $n=6$, $8$, $10$, $15$, $15'$, $24$ and $27$.
These can be constructed from the Polyakov loop in the fundamental representation using group theory as follows \cite{Gupta:2007ax}:
\begin{eqnarray}
L_6&=&\frac{1}{6}(l_3^2-l_3^*) \; , \label{l6} \\
L_8&=&\frac{1}{8}(|l_3|^2-1) \; , \\
L_{10}&=&\frac{1}{10}( l_3 \cdot l_6-l_8) \; , \\
L_{15}&=&\frac{1}{15}(l_3^* \cdot l_6-l_3) \; , \\
L_{15'}&=&\frac{1}{15}(l_3 \cdot l_{10}-l_{15}) \; , \\
L_{24}&=&\frac{1}{24}(l_3^* \cdot l_{10}-l_{6}) \; , \\
L_{27}&=&\frac{1}{27}(|l_6|^2-l_{8}-1) \; , \label{l27}
\end{eqnarray}
where $l_3=3 L_3^{\rm bare}$ and $l_3^*$ is its complex conjugate. From the above equation
it is clear that the Polyakov loops in all representations are normalized by the dimension
of the representation and thus will approach one at very high temperatures.
	
\section{Gradient flow renormalization}\label{sec:ren}

\begin{figure}
	\hspace*{-10mm}
	\includegraphics[width=0.45\textwidth,clip]{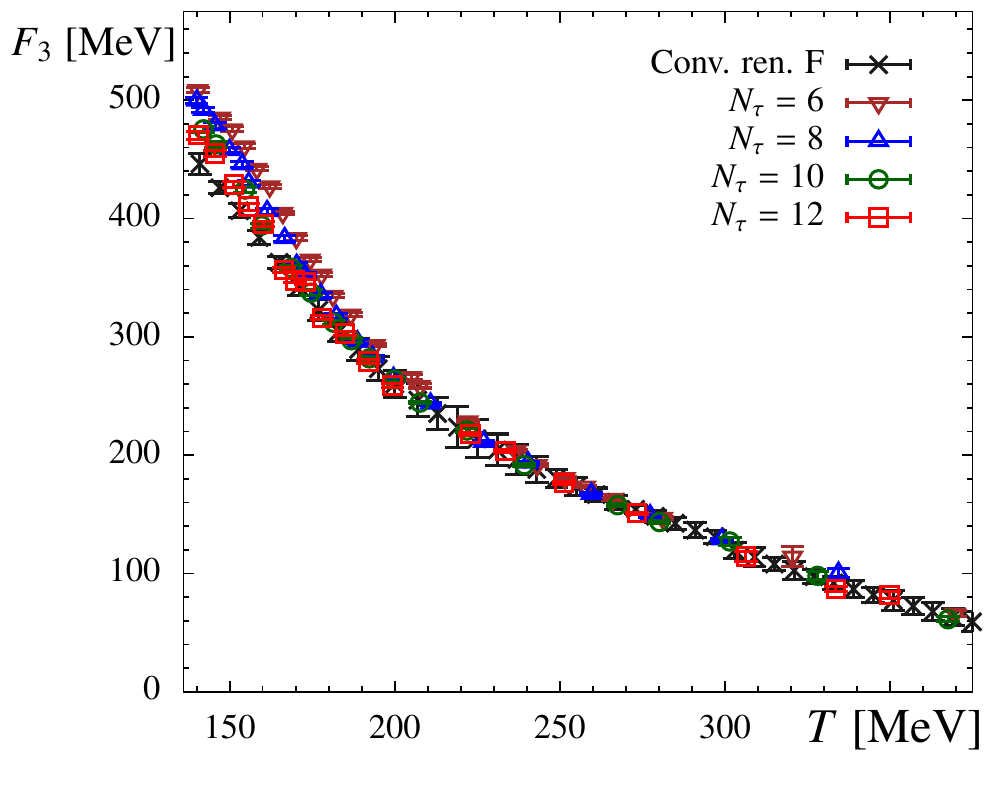}
	\caption{The free energy $F_3$ obtained from the gradient flow compared to the continuum results
         for $F_3$ in conventional renormalization scheme from Ref. \cite{new}.}
	\label{fig:fundfenergycomp}
\end{figure}

As discussed above, the Polyakov loop needs multiplicative renormalization.
We use the gradient flow to renormalize the expectation values of Polyakov loops.
The gradient flow is defined by the differential equation (equation of motion)
\cite{Luscher:2010iy}
\begin{equation}\label{eq:grad}
	\dot V_t(x,\mu) = -g_0^2(\partial_{x,\mu}S[V_t])V_t(x,\mu) \; ,
\end{equation}
where $g_0$ is the bare gauge coupling and $S[V_t]$ is the Yang-Mills action.
The field variables $V_t(x=(\mathbf{x},\tau),\mu)$ are defined on the four dimensional lattice
and satisfy the initial condition
\begin{equation}
	V_t(x,\mu)|_{t=0} = U_{\mu}(x) \; ,
\end{equation}
with $U_{\mu}(x=(\mathbf{x},\tau))$ being the usual SU(3) link variable and $t$ 
is a new index for the evolution in flow time and has dimension $[a^2]$. 
So far we have not specf.ed the discretization scheme for the Yang-Mills action. 
We could use the simple Wilson gauge action \cite{Luscher:2010iy}
or the tree-level improved Symanzik action \cite{Fodor:2014cpa} for $S[V_t]$. 
One usually refers to these schemes as Wilson- or Symanzik flow.
In our study we use the Symanzik flow, i.e. $S[V_t]$ is the tree-level improved Symanzik gauge action. We also performed some
calculations using Wilson flow, which are discussed in Appendix \ref{app:flowtype}. The differential equation is solved using a Runge-Kutta like 
scheme up to the desired value of $t$. For the flow evolution of the gauge configurations and for the calculation of the 
observable we use the MILC code \cite{MILC}.

Since Eq.~(\ref{eq:grad}) has the form of a diffusion equation, the gradient flow smears the original field $U_{\mu}(x)$
at the length scale 
\begin{equation}
f=\sqrt{8t} \; .
\end{equation}
For this reason operators evaluated at nonzero flow time, i.e. operators that are constructed from $V_t(x,\mu)|_{t>0}$ 
instead of $U(x,\mu)$ do not
require renormalization \cite{Luscher:2011bx}, the 
short distance singularities are removed by the gradient flow.
Furthermore, it can be shown that renormalized operators at $t=0$ are equal to
operators at nonzero flow time up to multiplicative constant \cite{Luscher:2011bx,Luscher:2013vga} if the flow time is
sufficiently small, $a \ll f \ll \Lambda_{\rm{QCD}}$. Therefore, the renormalization of the Polyakov loop can be achieved by
replacing the original link variables in Eq.~(\ref{eq:pl}) with  $V_t(x,\mu)|_{t>0}$. The choice of flow time $f=\sqrt{8t}$ corresponds
to a particular renormalization scheme as long as 
\begin{equation}
a \ll f \ll 1/T \; .
\label{eq:constr}
\end{equation}
To demonstrate the above point in Fig. \ref{fig1} we show the bare and the renormalized Polyakov loop constructed
from $V_t(x,\mu)|_{t>0}$ corresponding to the choice $f=f_0=0.2129$ fm \footnote{The value of $f_0$ corresponds
to the lattice spacing of the $32^3 \times 8$ lattices at the lowest temperature used in our analysis, and thus
it provides a natural unit for the flow time $f$.}.
In what follows we will give the flow time in units of $f_0$.
One can see that the strong $N_\tau$ dependence
of the bare loop is gone in the renormalized Polyakov loop as expected.

For the calculation of the renormalized Polyakov loop in an extended temperature region we need to change $f$ such that
the constraint given by Eq.~(\ref{eq:constr}) is always satisfied. To do so we proceed as follows: 
We define regions where the flow time $f$ is constant in physical units, 
which means that changing the temperature $T$ via the lattice spacing means changing the flow time $t$ in the 
actual calculation such that $f =\rm{const}$ in fm. Different choices of $f$ correspond to different renormalization
schemes. For this reason
the free energy should be independent of the flow time up to a constant shift, i.e. $F_n(f)-F_n(f')$ is approximately $T$-independent. 
We are limited in the range of $N_\tau$ and therefore, as we want to cover a very broad temperature range, 
we have to define different flow regions to fulfill this condition:
\begin{equation}\label{eq:flowdef}
	f = \left\{ \begin{array}{rl}
	3 f_0 & \; \mbox{for} \;\; T \, < \, 200~{\rm MeV} \; ,\\
	2 f_0 & \; \mbox{for} \;\; 200~{\rm MeV} \leq T \, \leq \, 300~{\rm MeV} \; ,\\
	0.50 f_0 & \; \mbox{for} \;\; 300~{\rm MeV} \leq T \, < \, 600~{\rm MeV} \; ,\\
	0.25 f_0 & \; \mbox{for} \;\; T \, \geq \, 600~{\rm MeV} \; ,\\
	\end{array} \right.
\end{equation}
where $f_0=0.2129$ fm.
The different regions can then be matched by a constant shift of the free energy and we do this by determining the shift via an 
overlapping temperature point between the flow regions for the different ensembles. We would like to compare the 
renormalized Polyakov loop obtained
with gradient flow to the conventional renormalization of the Polyakov loop based on the static potential at zero temperature.
We use the continuum extrapolated results for the renormalized Polyakov loop obtained using the normalization condition
of the potential $r_1 V(r=r_1)=0.2605$ \cite{new}. So we need to match the gradient flow scheme with the potential based
conventional renormalization scheme.
This is done by matching the values of the free energy at a single temperature point for $N_\tau=12$ ($T\approx 200$~MeV). 
The other ensembles will be shifted by the same amount, which guarantees that the cutoff effects from the different $N_\tau$ are not obscured.

After performing this shift we show in Fig.~\ref{fig:fundfenergycomp} the free energy of the static charge $F_3=-T\ln P_3$ in the fundamental
representation for different $N_\tau$. At low temperatures, $T<200$ MeV, we see some $N_\tau$ dependence and the free energy obtained
from the gradient flow is larger, but approaches the continuum result with increasing $N_\tau$. The largest deviations from
the continuum results are about $10\%$ for $N_\tau=6$ and are few percent for $N_\tau=12$.
For $T>200$ MeV the cutoff effects
are much smaller and we see agreement with free energy obtained using conventional renormalization and the results
for the free energy obtained using gradient flow. Clearly any difference between the two approaches should vanish
in the continuum limit. We performed the continuum extrapolation of the free energy for different values of the flow time
and verified that this is indeed the case. 
In particular, the deviations between the results obtained in the two renormalization schemes that can be seen in Fig. \ref{fig:fundfenergycomp}
disappear after taking the continuum limit.
The details of this analysis are presented in Appendix \ref{app:cont}, where we
also show explicitly that different choices of the flow time amount to a constant shift in $F_3$.

We close this section by noting that we also calculated the Polyakov loop in fundamental representation for
the smaller light quark masses, namely $m_l=m_s/40$. Compared to the $m_l=m_s/20$ results we see a downward 
shift of $F_3$. This shift in the free energy is consistent with the shift in the deconfining temperature
of about $3$ MeV, i.e. by shifting the $m_s/40$ data by $3$ MeV to larger temperatures we make them agree
with the $m_s/20$ data. 
The mass dependence of $F_3$ is discussed in detail in Appendix \ref{app:massdep}.
Next we want to use the gradient flow approach to calculate the renormalized Polyakov loop in higher representations.

\section{Polyakov loop in higher representations}

\begin{figure}
\includegraphics[width=8cm]{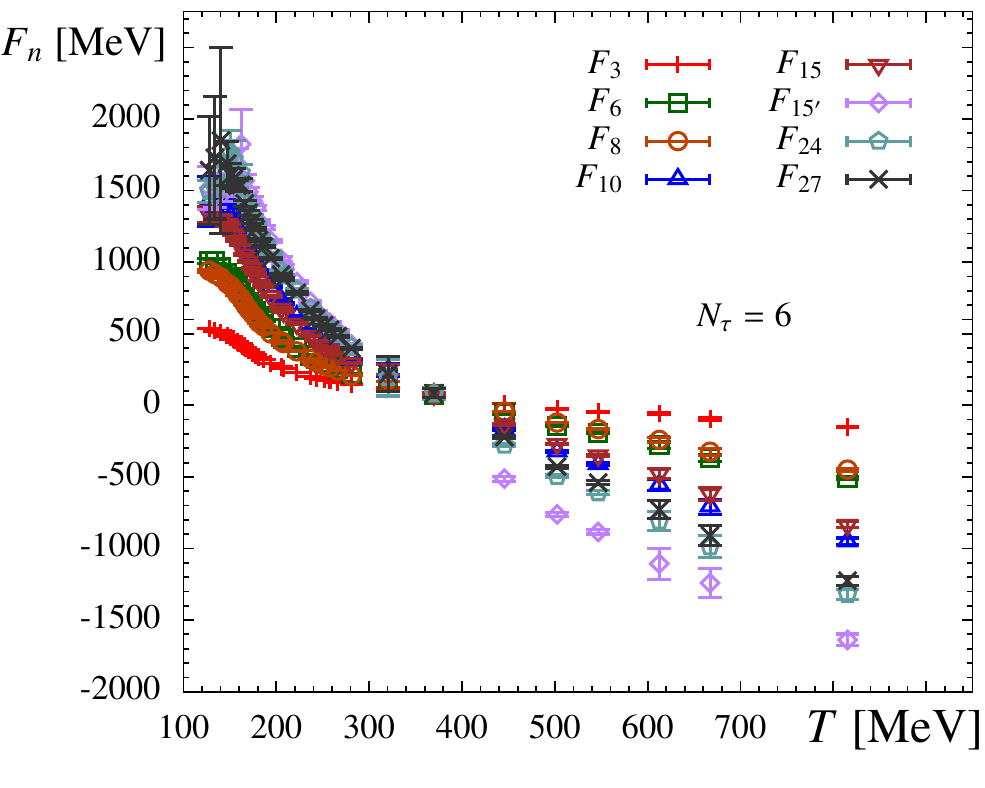}
\caption{The free energy of the static charge in different representations as function of the
temperature for $N_\tau=6$.}
\label{fig:higherreps}
\end{figure}

We calculated the expectation value of the Polyakov loop in higher representations, namely
sextet, octet, decuplet, $15$, $15'$, $24$ and $27$ using Eqs.~(\ref{l6})-(\ref{l27}) and the gradient
flow in the same manner as described in the previous section. In particular, we used the flow times
defined by Eq.~(\ref{eq:flowdef}) also here. 
The numerical results for the Polyakov loops in higher representations for $N_\tau=6$ are shown in terms
of  the corresponding free energies
$F_n=-T\ln P_n$ in Fig. \ref{fig:higherreps}. The vertical scale in the figure has
been shifted by $100$ MeV so that the value of $F_3$ is the same as in the previous section.
For representation $15'$ the data for flow time $2 f_0$
are very noisy around $T=300$ MeV and are therefore not shown in the figure.
The free energy of the static charge is larger for
the higher representations at low temperatures. This can be understood as follows:
At very low temperatures the free energy of the static charge is determined by the binding energy
of the lightest static-light hadron that can screen that charge. For the free energy in the fundamental
representation it is given by the mass of the lightest static-light meson, for the free energy in
the sextet representation it is given by the binding of baryon with two static and one light quarks, for
the free energy in the octet representation it is determined by the gluelump mass, while for 
higher representation it is determined by binding energies of more exotic states. 
The free energies follow the hierarchy that one expects for the hierarchy of the binding energies of
the corresponding hadrons, e.g. the binding energy of static-light meson can be estimated to be
around $600$ MeV, the binding energy of static-light baryon to be around $1$ GeV \cite{Wagner:2011fs}, while the gluelump
binding energy is about $2$ GeV \cite{Simonov:2000ky}.
The larger values of the free energies at low temperature lead to smaller signals and thus 
more noisy data. In fact, without the gradient flow it is impossible to extract signals for the Polyakov loops
in higher representations. The gradient flow increases the signal by removing the ultraviolet noise if
the flow time is sufficiently large. Large flow times result in better signal.
Unlike for the fundamental Polyakov loops extracting signals at low
temperatures with flow time smaller than defined by Eq.~(\ref{eq:flowdef}) is challenging.

\begin{table}[b]
	\caption{ 
	Values of the quadratic Casimir $C_n$ in representation $n$ and its ratio to 
the value of the fundamental quadratic Casimir $R_n=C_n/C_3$. See, e.g., \cite{Gupta:2007ax} for details.
	}
	\label{tab:casimirstable}
	\begin{center}
		\vspace{-0.0cm}
		\begin{tabular}{m{0.12\textwidth}m{0.12\textwidth}m{0.12\textwidth}}
			\hline \hline
			$n$ & $C_n$ & $R_n$ \\
			\hline
			$3$ & 4/3 & 1 \\
			$6$ & 10/3 & 5/2 \\
			$8$ & 3 & 9/4 \\
			$10$ & 6 & 9/2 \\
			$15$ & 16/3 & 4 \\
			$15'$ & 28/3 & 7 \\
			$24$ & 25/3 & 25/4 \\
			$27$ & 8 & 6\\
      \hline
		\end{tabular}
	\end{center}
\end{table}

The temperature dependence of the free energy of static charges is larger for higher representations
as can be seen in Fig. \ref{fig:higherreps}. At highest temperatures the free energies in the higher
representations are negative and are larger in absolute value than the free energy in the fundamental
representation. This also means that Polyakov loops in higher representations are significantly larger
than one. These features can be understood in term of the weak coupling calculations.
In leading order perturbation theory the free energy of static charges is 
\begin{equation}
F_n(T)=-C_n \alpha_s m_D,
\end{equation}
where $C_n$ is the quadratic Casimir operator of representation $n$, $\alpha_s$ is the coupling constant, and
$m_D \sim \sqrt{4 \pi \alpha_s} T$ is the leading order Debye mass. 
The values of the quadratic Casimir operators are given in Table \ref{tab:casimirstable}.
According to the above
equation the free energy of static charges satisfies Casimir scaling, i.e. the free energies in various
representations only differ by the value of $C_n$. This Casimir scaling holds in perturbation theory
up to order $\alpha_s^3$ \cite{Brambilla:2010xn}\footnote{To see this it is important
to re-exponentiate the perturbative expansion of $P_n$ in terms of $F_n$.}.
In terms of the Polyakov loops the Casimir scaling implies
\begin{equation}
P_3=P_6^{1/R_6}=P_8^{1/R_8}=P_{10}^{1/R_{10}}=... \; ,
\label{scaling}
\end{equation}
where $R_n=C_n/C_3$. The values of $R_n$ are also given in Table \ref{tab:casimirstable}.

Non-perturbatively Casimir scaling of the Polyakov loop was studied on
the lattice in SU(N) gauge theories as well as in two-flavor QCD with heavy quarks.
In these studies the renormalized Polyakov loop in higher representations was calculated
assuming the Casimir scaling for the renormalization constants of the Polyakov loop, i.e.
by rescaling the renormalization constants of the fundamental representation
$Z_n=Z_3^{R_n}$.  This assumption is closely related to the Casimir
scaling of the zero temperature potentials since the renormalization constants
are related to the potentials.
In SU(3) gauge theory the zero temperature potentials between
static charges in various representations have been calculated \cite{Bali:2000un}.
It has also been shown that Casimir scaling holds 
for the potentials after subtracting the UV divergent
part from the potentials to an accuracy better that $5\%$ 
for distances $r<1$~fm~\cite{Bali:2000un}.
Furthermore, Casimir scaling of the zero temperature 
potentials holds up to order
$\alpha_s^4$ in perturbation theory and its breaking is numerically small \cite{Anzai:2010td}.
The Casimir scaling of the potential in SU(3) gauge theory
is of course only approximate. For large
enough distances it is clearly violated since the potential in the adjoint
and higher representations will saturate at some finite value of $r$ due to 
string breaking, while the potential in the fundamental representation is
linearly rising with $r$. In general, 
Casimir scaling does not hold for the phenomenon of string breaking;
string breaking in various representations will happen at different
distances determined by the masses of various static hadrons (see the discussion above).
Strictly speaking the renormalization of the Polyakov loop in higher
representations would require calculating and fixing the normalization $e_n(g_0)$
of the potentials in different representations at zero temperature, i.e.
one needs to define separate renormalization constants for each representation
independently. The choice $Z_n=Z_3^{R_n}$ is just one economical
scheme for defining the renormalization constants in higher representations.

Using the gradient flow we can calculate the renormalized Polyakov loop
in higher representations without any assumptions. Furthermore, 
as already discussed the gradient flow is instrumental for obtaining signals
for the Polyakov loops in higher representations at low temperatures. In fact,
we are not aware of any other methods that can achieve this in full QCD.

\begin{figure*}
\includegraphics[width=5.8cm]{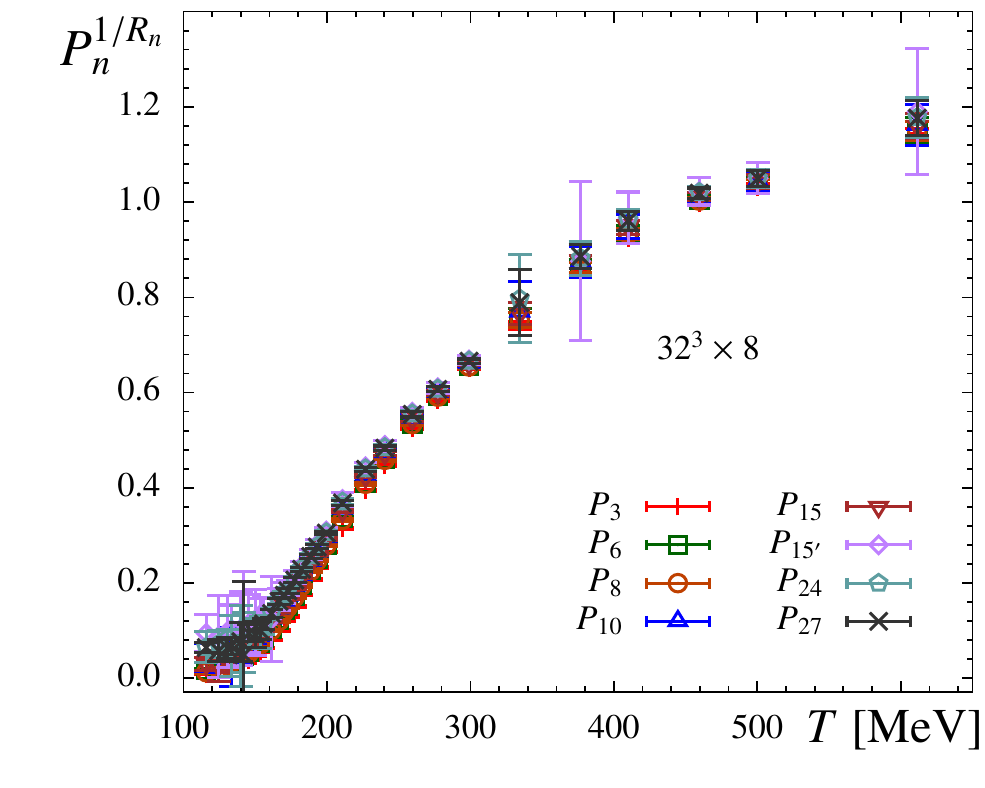}
\includegraphics[width=5.8cm]{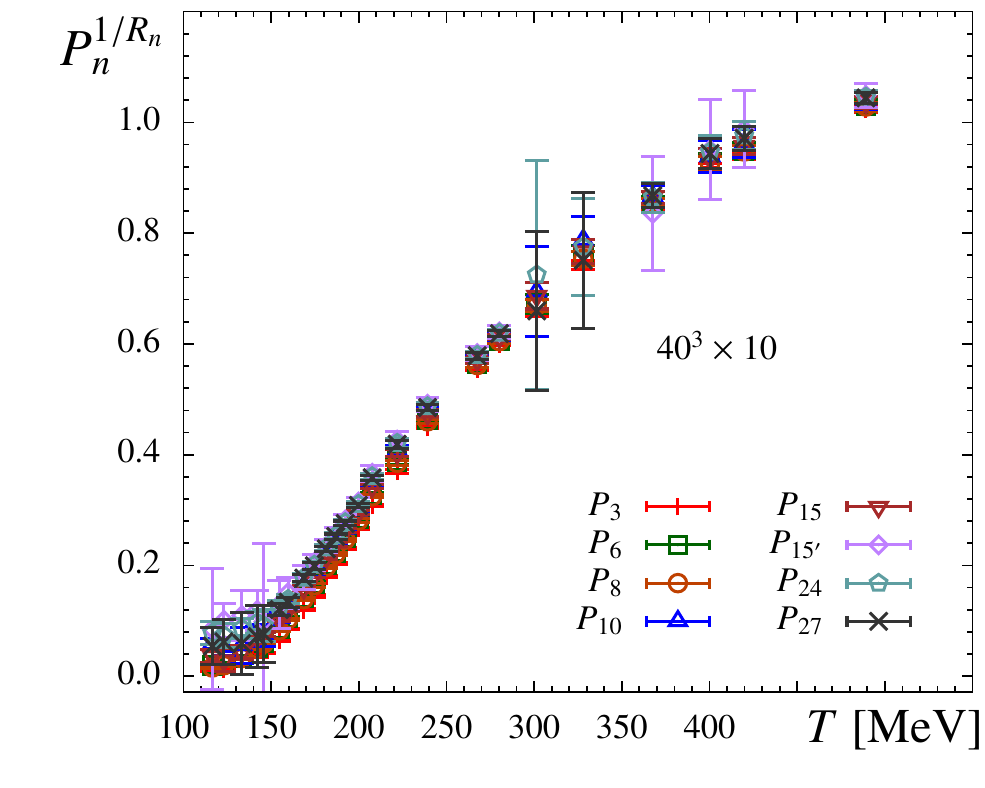}
\includegraphics[width=5.8cm]{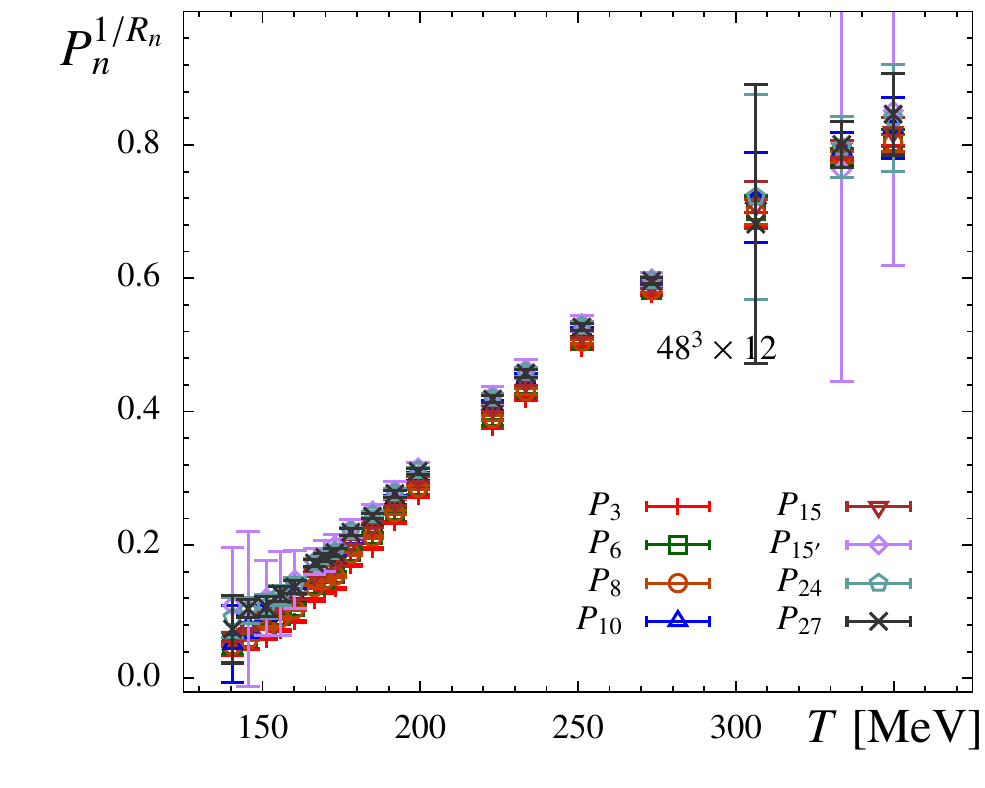}
\caption{Polyakov loops in various representations scaled by the ratio of the
appropriate Casimir operators (see text) for $N_\tau=8$ (left), $N_\tau=10$ (middle) and $N_\tau=12$ (right).
The Polyakov loops have been rescaled by $\exp(-\Delta_3/T)$, $\Delta_3=100$ MeV,
to match to the conventional scheme for the fundamental Polyakov loop.}
\label{fig:scaling} 
\end{figure*}

In Fig. \ref{fig:scaling} we show the renormalized Polyakov loops in various 
representations for $N_\tau=8,~10$ and $12$ scaled by the ratio of the corresponding
Casimirs (cf. Eq.~(\ref{scaling})). Here we also impose the additional normalization
that connects the free energy in the conventional renormalization scheme and
the gradient flow renormalization scheme. At high temperatures we observe Casimir
scaling. This is expected based on the previous lattice
studies. At low temperatures, on the other hand,  we see deviations from the Casimir scaling.
From the above figure one can also see that the $N_{\tau}$ dependence of the Polyakov loop
is small. In fact, no $N_{\tau}$ dependence is seen within errors in Fig. \ref{fig:scaling}.
Thus cutoff effects in the Polyakov loop in higher representations are under control.

\begin{figure*}
\includegraphics[width=8cm]{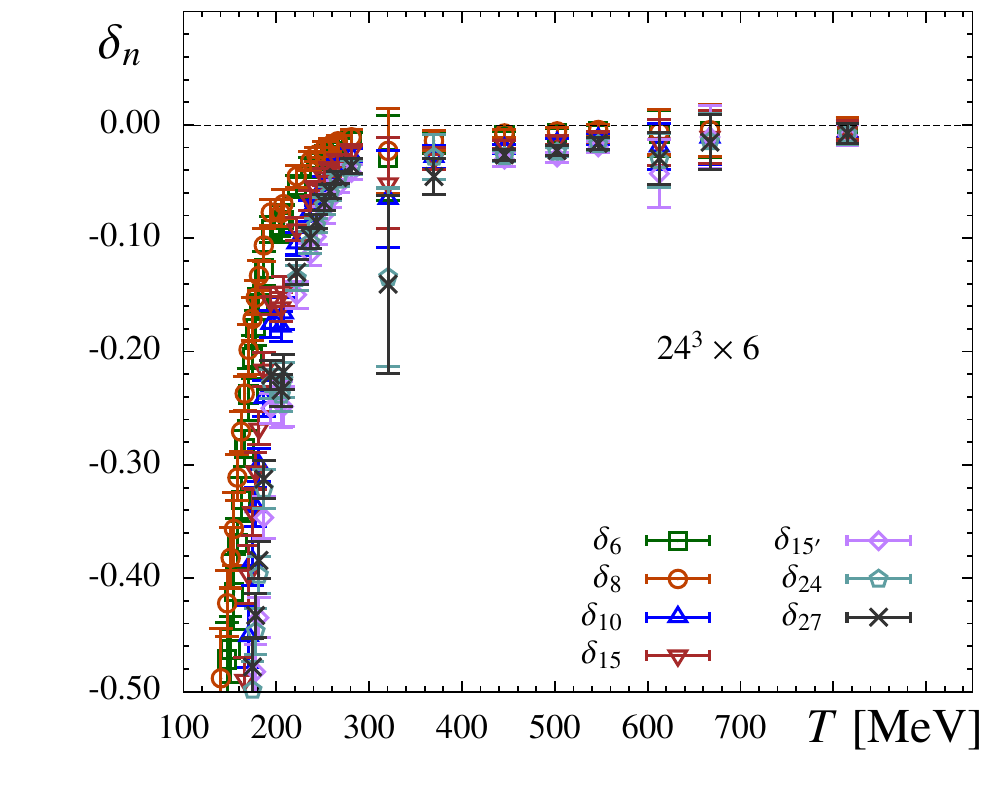}
\includegraphics[width=8cm]{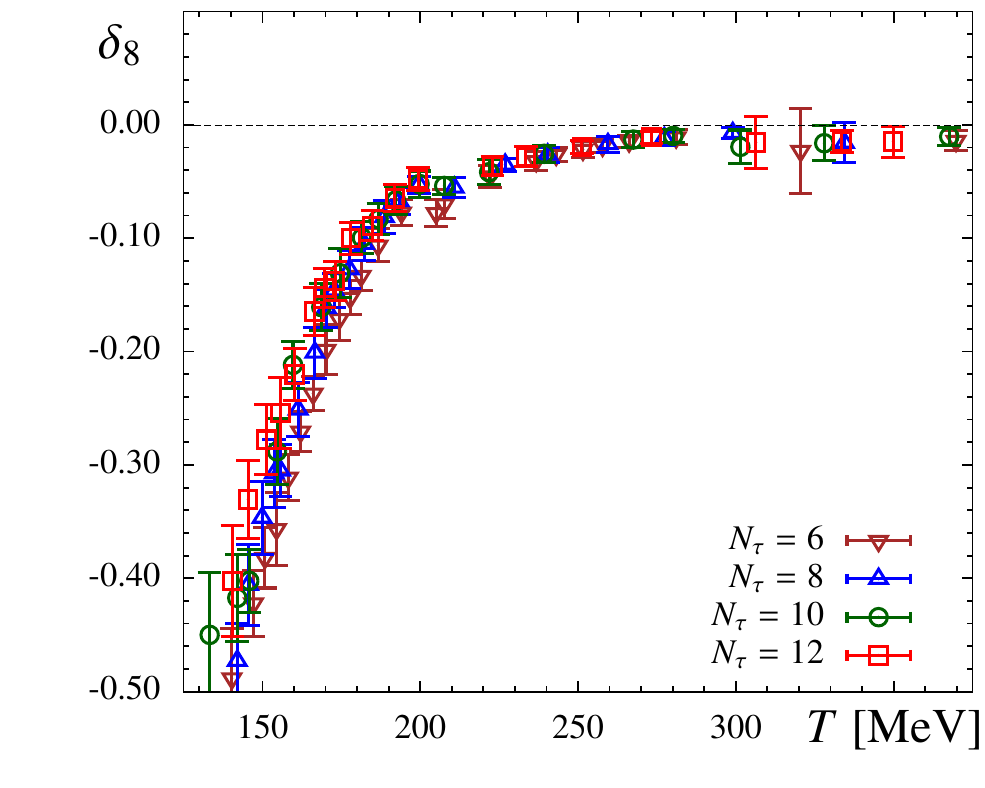}
\caption{The ratios $\delta_n$ characterizing the breaking of Casimir scaling 
for $N_\tau=6$ (left) and various representations $n$, and for the octet representation and various $N_\tau$ (right).}
\label{fig:breaking}
\end{figure*}

The deviations from Casimir scaling at low temperatures cannot be seen well
in Fig. \ref{fig:scaling} since the Polyakov loops are small there.
To clearly see deviations from the Casimir scaling at low temperatures
we introduce the dimensionless combination
\begin{equation}
\delta_n=1-\frac{P_n^{1/R_n}}{P_3}.
\end{equation}
Our numerical results for $\delta_n$ for various representations and various $N_\tau$ are
shown in Fig. \ref{fig:breaking} as a function of the temperature. 
We see that breaking of Casimir scaling for $T>250$ MeV
is of the order of a few percent, but becomes significant for lower temperatures. 
This is the first time that breaking of Casimir scaling for the Polyakov loop is
seen in lattice calculations. In previous studies no conclusive statements could
be made due to large statistical errors or large volume effects. 
The above results imply that for $T>250$ MeV color screening follows the perturbative pattern,
while at lower temperatures it is strongly nonperturbative.
Figure \ref{fig:breaking} also shows that the breaking of Casimir scaling is independent
of the value of $N_\tau$. We demonstrate this in the case of the octet representation.
This is another way to see that cutoff effects in the higher
representations are small. 

One may ask to what extent the observed Casimir scaling or its breaking depends
on the value of the flow time. Different flow time corresponds to different renormalization
schemes, i.e. to different choices of $Z_n$. We calculated $\delta_n$ for flow times
$f=f_0$, $2 f_0$ and $3 f_0$ and we do not see significant flow time dependence of this
quantity. The corresponding numerical results are presented in Appendix \ref{app:flowtime}.
Therefore, the above statements about the breaking of the Casimir scaling at
low temperatures are independent on the choice of $f$.

We also examined the volume dependence of the Polyakov loop
in higher representations and did not find significant volume dependence. 
Thus, the observed breaking of the Casimir scaling is not a finite volume effect.
The details of this analysis are given in Appendix \ref{app:massdep}.

\section{Summary and discussion}

We discussed the renormalization of the Polyakov loop with the gradient flow. 
We applied the gradient flow with the Symanzik gauge action, i.e. the Symanzik flow, and calculated 
the Polyakov loop after the evolution of the gauge fields in flow time up to a fixed
value $f=\sqrt{8t}$ in physical units, which fixes the renormalization scheme for the free energy. 
With this approach it was possible to cover a wide temperature range from temperatures as low as $T=116$~MeV 
and up to $T=815$~MeV. We compared our results for the fundamental Polyakov loop $P_3$ with results 
for the renormalized Polyakov loop obtained in the conventional scheme based on the static potential, 
and found very good agreement at all temperatures. 

In addition we calculated the renormalized Polyakov loop in higher representations.  We found that Casimir scaling 
is approximately fulfilled for full QCD for temperatures above $T=250$~MeV in agreement with
previous studies,  possibly indicating the weakly coupled nature of quark gluon plasma at high temperatures.
At lower temperatures, however, we found for the first time large deviations from Casimir scaling. 

The renormalization of the Polyakov loop with gradient flow is very useful for studying its behavior at high
temperatures, where performing zero temperature calculations is very costly. We will discuss this in
a forthcoming publication \cite{new}.

\vspace{-0.5cm}

\acknowledgments{
This work was supported by U.S. Department of Energy under Contract No. DE-SC0012704. H.-P. Schadler was funded by the FWF DK W1203, ``Hadrons in Vacuum, Nuclei and Stars". The authors want to thank Johannes Weber for interesting discussions. The numerical computations have been carried out on the clusters of USQCD collaboration, on the Vienna Scientific Cluster (VSC) and in NERSC with the publicly available MILC code.
}

\appendix
\section{COMPARISON OF SYMANZIK AND WILSON FLOW}
\label{app:flowtype}

\begin{figure}[t]
	\centering
	\hspace*{-10mm}
	\includegraphics[width=0.42\textwidth,clip]{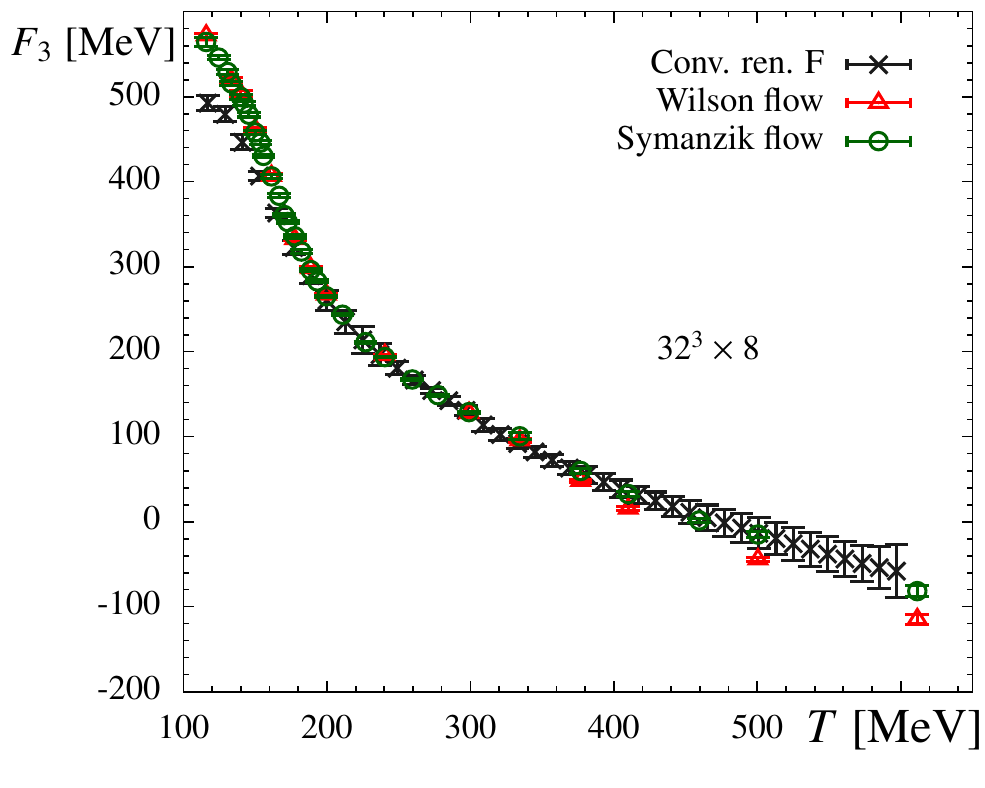}
	\caption{We compare the renormalized fundamental Polyakov loop free energy $F_3$ obtained with Wilson and Symanzik flow as described in Sec.~\ref{sec:ren}. In addition we show the continuum extrapolated, renormalized free energy from \cite{new} (black triangles). While for lower temperatures the cutoff effects are small, at higher temperatures the Wilson flow shows larger deviations from the conventional result than the Symanzik flow.}
	\label{fig:wilsonflow}
\end{figure}

In this appendix we compare the renormalized fundamental Polyakov loop obtained with the Symanzik flow,
as described in Sec.~\ref{sec:ren},
with the renormalized fundamental Polyakov loop obtained with the same procedure but using 
the Wilson gauge action in Eq.~(\ref{eq:grad}), i.e. using the Wilson flow. 
In Fig.~\ref{fig:wilsonflow} we show $F_3$ obtained using Symanzik flow and Wilson flow,
and compare these with the conventionally renormalized free energy. 
We use the same flow times and same matching procedure for the Wilson flow that we used for
the Symanzik flow (cf. Sec. III). 
From Fig. \ref{fig:wilsonflow} one finds 
that at low temperatures the cutoff effects are small and both results agree. 
At higher temperatures this changes as for temperatures above $350$~MeV the Wilson 
flow produces smaller values for the free energies and at some point it is below the results obtained
using the conventional renormalization procedure. This is most likely due to larger cutoff effects
in the case of the Wilson flow.

\section{FLOW TIME DEPENDENCE AND THE CONTINUUM LIMIT OF THE FREE ENERGY}
\label{app:cont}
\begin{figure*}
\includegraphics[width=6.5cm]{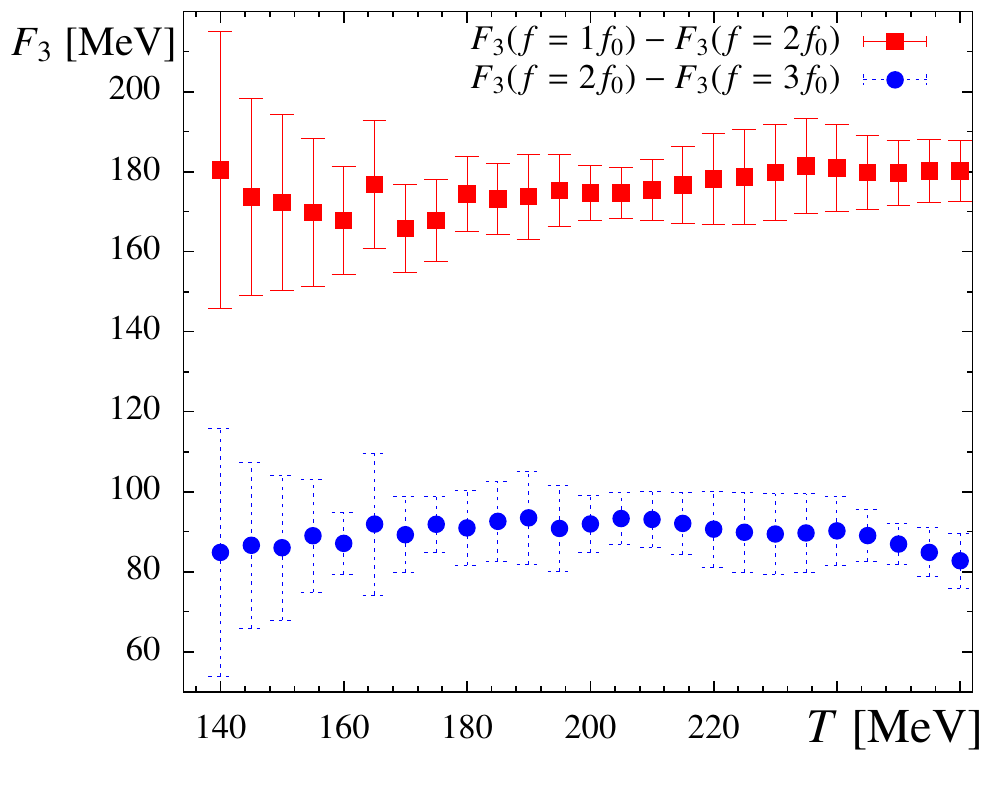} 
\includegraphics[width=6.5cm]{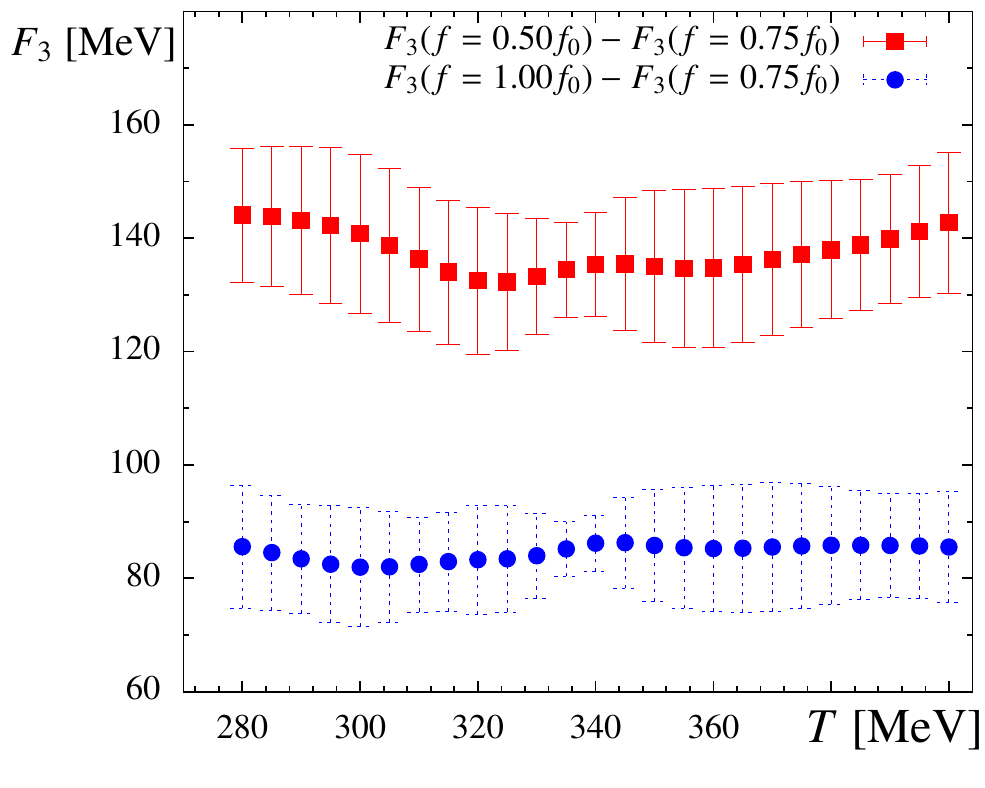}
\includegraphics[width=6.5cm]{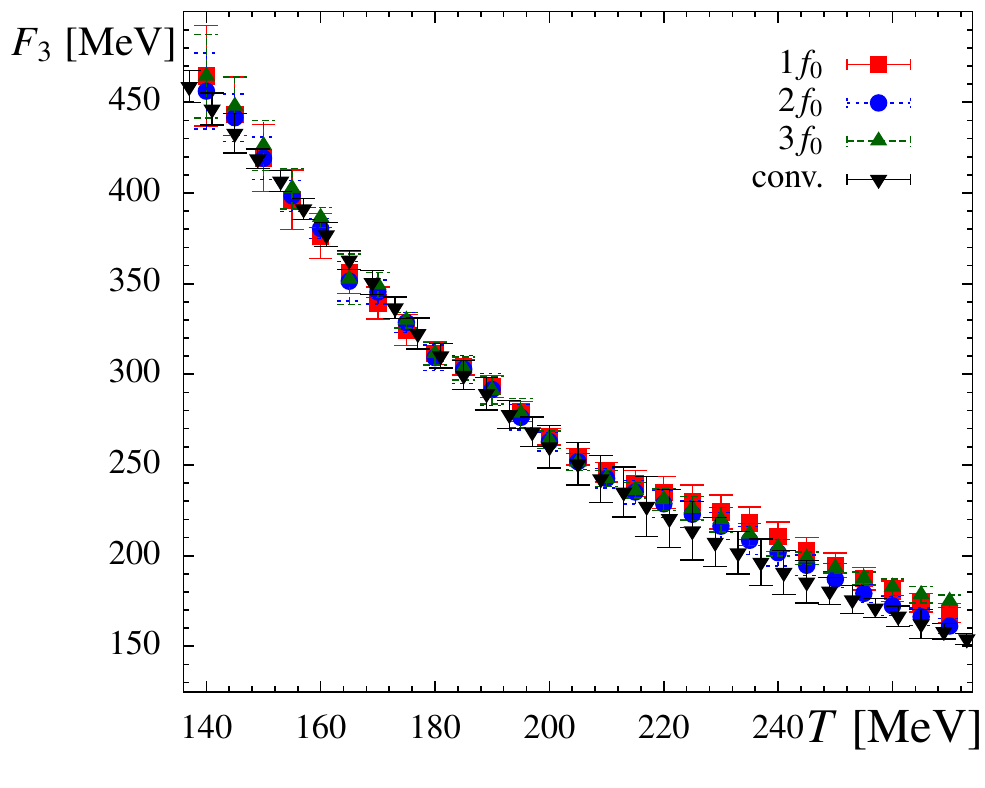}
\includegraphics[width=6.5cm]{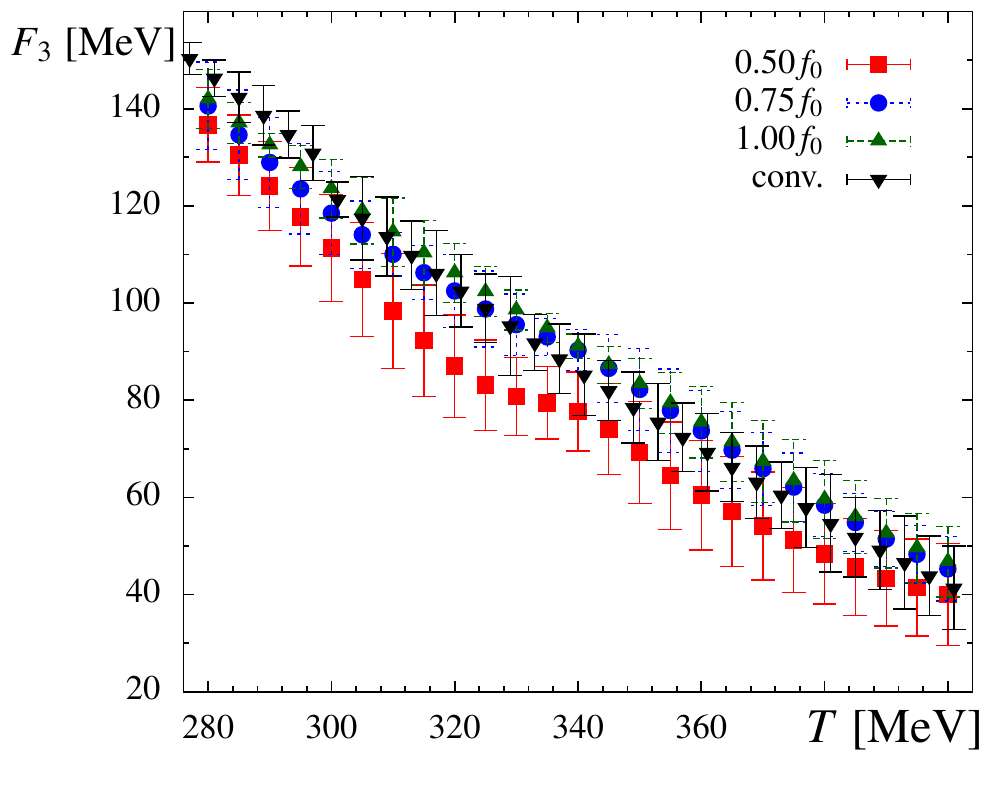}
\caption{The difference $F_3(f)-F_3(f')$ for different flow time (upper panels) and
the comparison of $F_3(f)$ with the continuum results in the conventional scheme (lower panels).
The results in the low and high temperature regions are shown separately
in the left and right panels, respectively.}
\label{fig:cont}
\end{figure*}

In this appendix we discuss the flow time dependence and the
continuum limit of the free energy of fundamental charge. 
As discussed in the main text in the continuum limit the free energy
of a static charge should be independent of the flow time up to
an additive temperature independent constant, i.e. $F_n(f)-F_n(f')$
should be temperature independent. Here we show that this
is indeed the case using the fundamental free energy as
an example. 
To perform the continuum extrapolations we 
split the temperature region used in our study into the low
temperature region, corresponding to $T<280$ MeV, and the high 
temperature region, corresponding to $T \ge 280$ MeV. 
In these intervals we use the values of $f$ that satisfy
the condition given by Eq.~(\ref{eq:constr}).
We perform interpolations of the free energy separately in 
these intervals using smoothing splines and the R package \cite{Rpackage}.
The errors of the interpolations are estimated by the bootstrap
method and in some cases adjusted such that they are
comparable to the statistical errors of the lattice data.
We perform continuum extrapolations  at temperatures separated by $5$ MeV using the form
$a+b/N_\tau^2+c/N_\tau^4$ and the results of these extrapolations are
shown in Fig. \ref{fig:cont}. As one can see from
the figure, after continuum extrapolations 
$F_3(f)-F_3(f')$ is temperature independent as expected.
In the figure we also compare our results for $F_3$
with the continuum extrapolated results obtained in
the conventional way. The continuum extrapolated results
obtained for different flow times have been shifted by
a constant to match the free energy in the conventional scheme.
After this shift our results agree with the results
obtained in the conventional scheme, in particular there is
no discrepancy at low temperature previously observed at 
fixed $N_\tau$ (cf. Fig. \ref{fig:fundfenergycomp}).

\section{QUARK MASS DEPENDENCE AND VOLUME DEPENDENCE OF THE RENORMALIZED POLYAKOV LOOPS}
\label{app:massdep}
\begin{figure*}
\includegraphics[width=6.5cm]{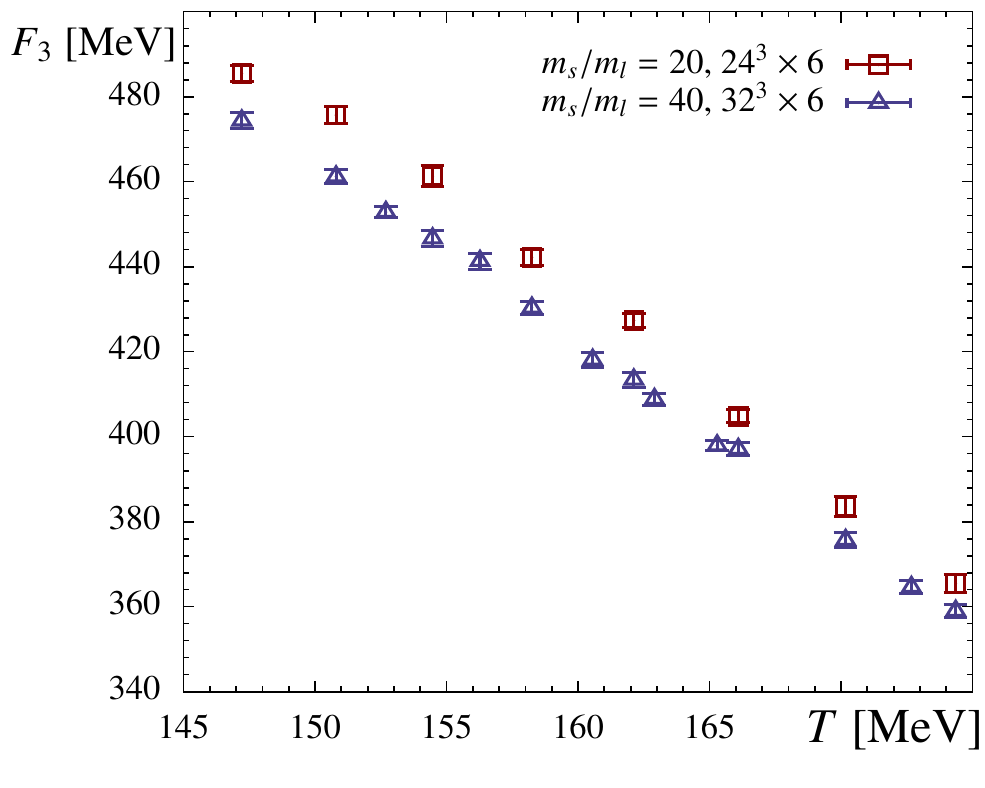}
\includegraphics[width=6.5cm]{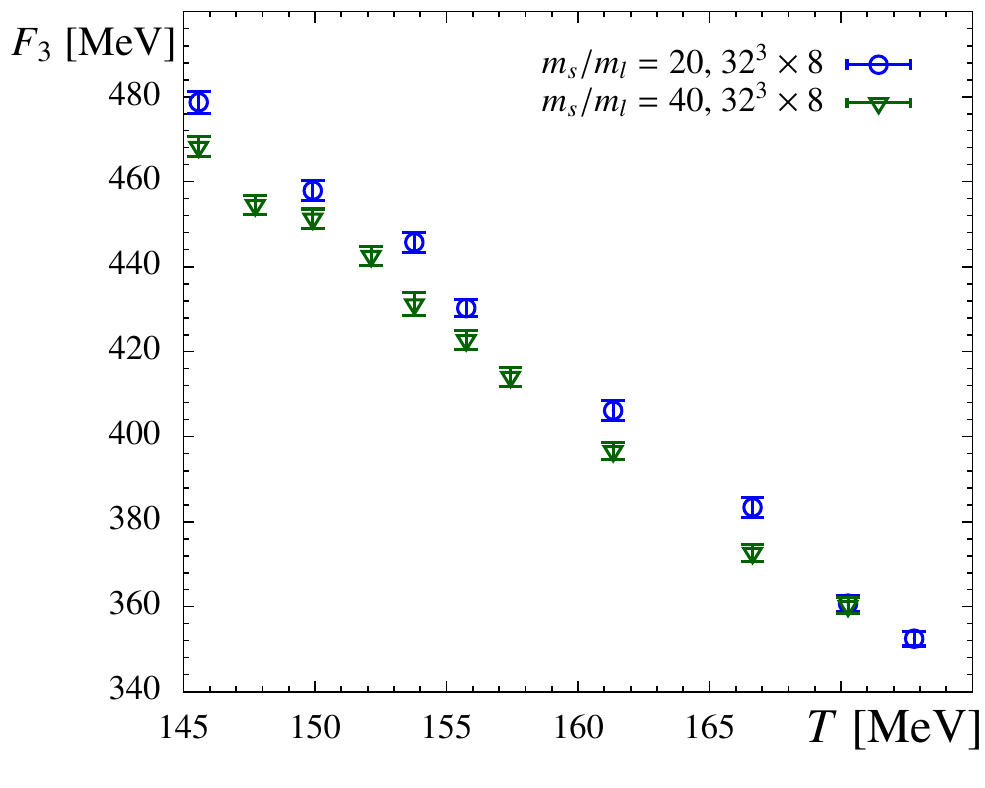}
\includegraphics[width=6.5cm]{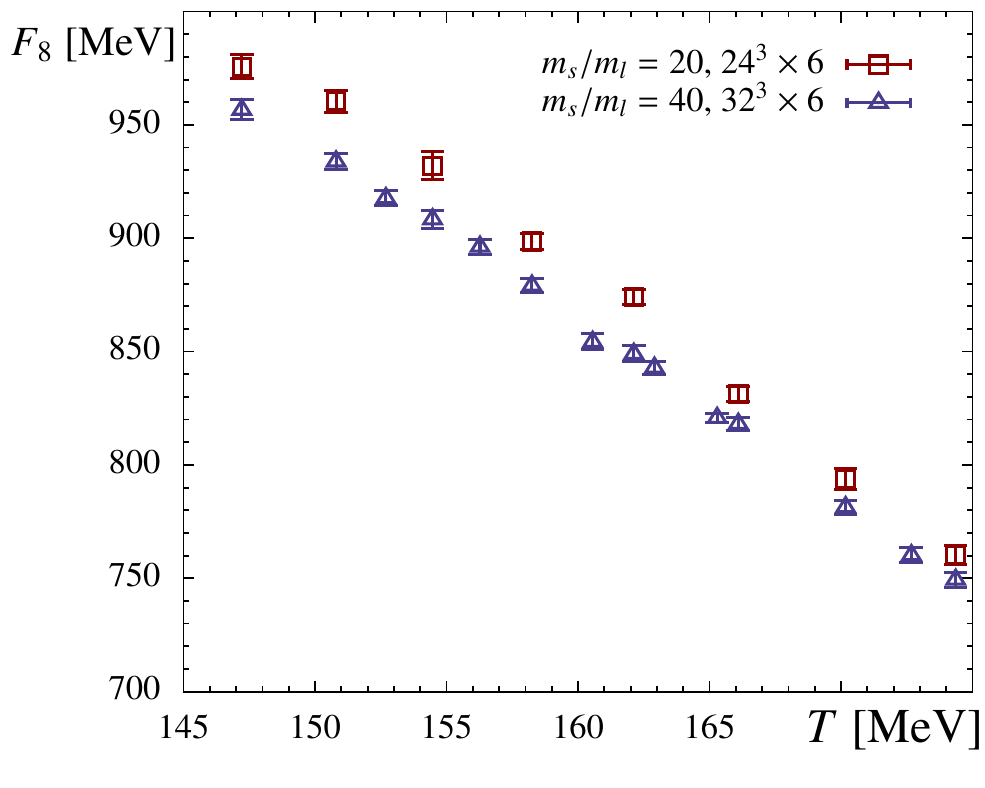}
\includegraphics[width=6.5cm]{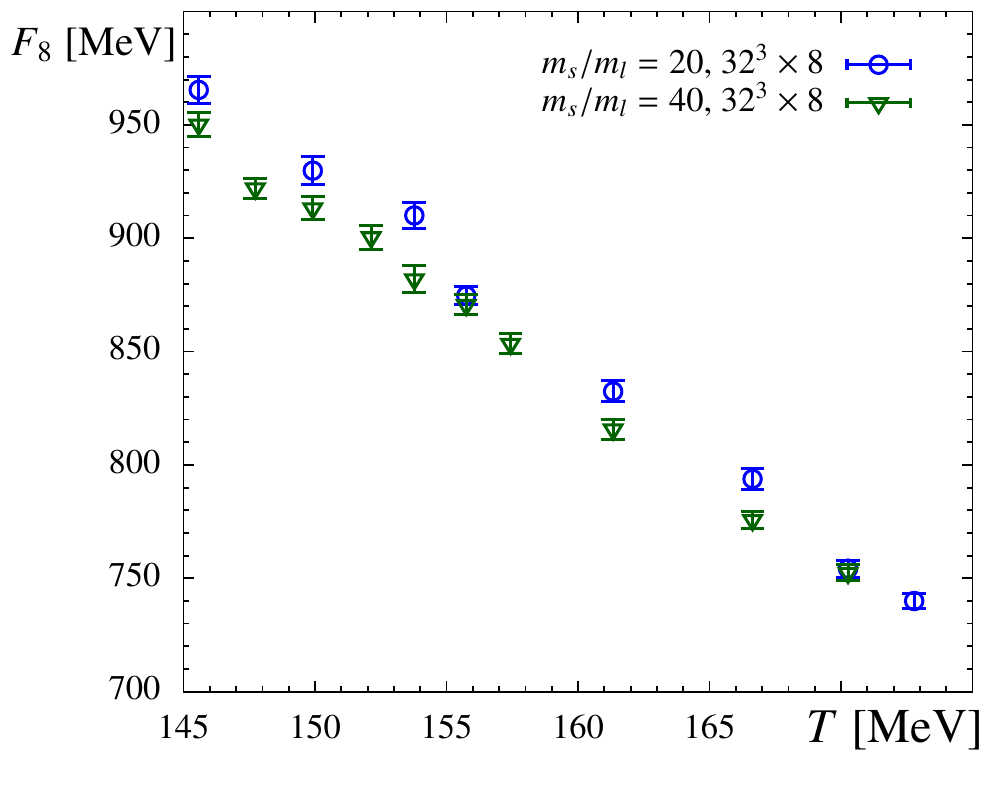}
\caption{The free energy in the fundamental (triplet) and adjoint (octet)
representations calculated for two different quark masses $m_l/m_s=1/20$
and $m_l/m_s=1/40$. The value $3 f_0$ is used for the flow time. 
The upper panels show the fundamental free energy and lower panels show
the adjoint free energy.}
\label{fig:massdep} 
\end{figure*}

In this appendix we discuss the quark mass and volume dependence
of the free energies $F_n$. In addition to the calculations
of the Polyakov loop for $m_l/m_s=1/20$ we also performed calculations for the smaller light
quark mass $m_l/m_s=1/40$ on $32^3 \times 6$ and $32^3 \times 8$ lattices. Since for $N_\tau=6$
we have two different volumes we can make some statements about finite volume effects as well.
In Fig. \ref{fig:massdep} we show the temperature dependence of the fundamental
and adjoint free energies as function of the temperature for two different
quark masses and flow time $3 f_0$. 
We see that the free energies show some quark mass dependence,
namely they are smaller for the smaller quark mass. The relative difference
of the free energies calculated for the two quark masses is about
the same for triplet and octet charges and for $N_\tau=6$ and $N_\tau=8$.
This difference may be understood in terms of change in the transition temperature. 
Shifting the $m_s/20$ data by $3$ MeV to lower temperatures almost eliminates this
difference.

For $N_\tau=8$ the spatial volume is the same for both quark masses,
but for $N_\tau=6$ the spatial volumes are different, namely we use
$24^3 \times 6$ and $32^3 \times 6$ volumes.
Since the shift
in the free energies is the same for $N_\tau=8$ and $N_\tau=6$ and can be understood
as a quark mass effect we conclude that volume effects in the free energies
in the fundamental and adjoint representations
are smaller than the estimated errors and thus can be neglected. Similar
conclusions can be made for the free energies in other representations.
Therefore the observed breaking of Casimir scaling is not affected by
finite volume effects.

\section{FLOW TIME DEPENDENCE OF CASIMIR SCALING}
\label{app:flowtime}

\begin{figure*}
\includegraphics[width=5.5cm]{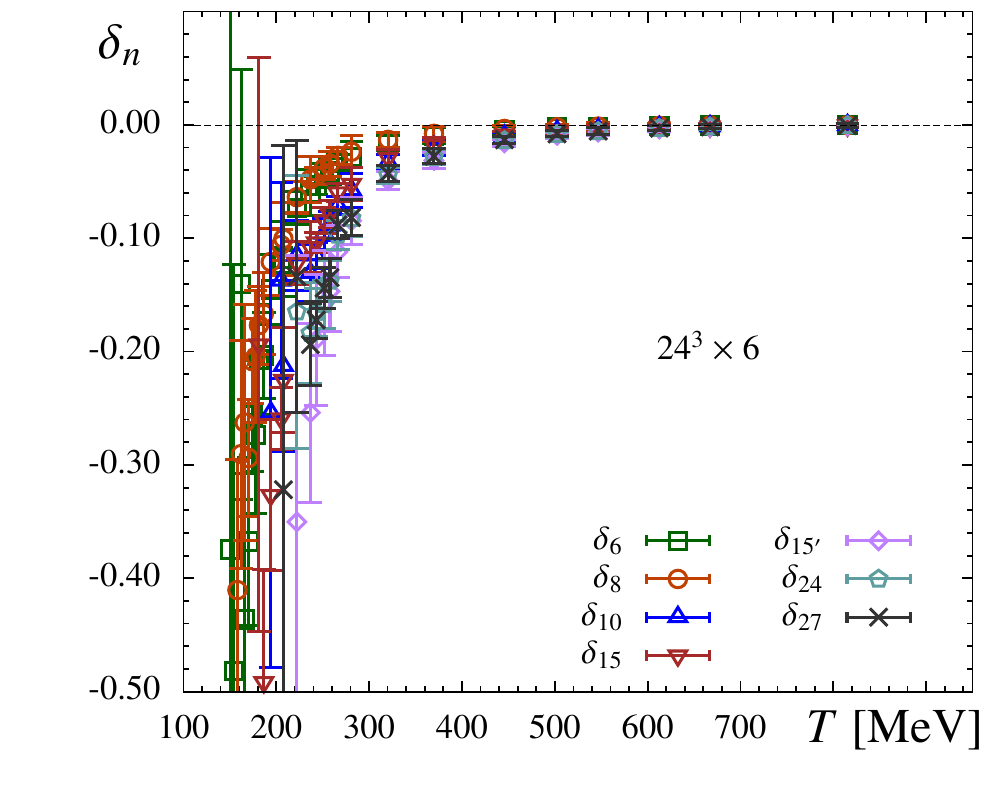}  
\includegraphics[width=5.5cm]{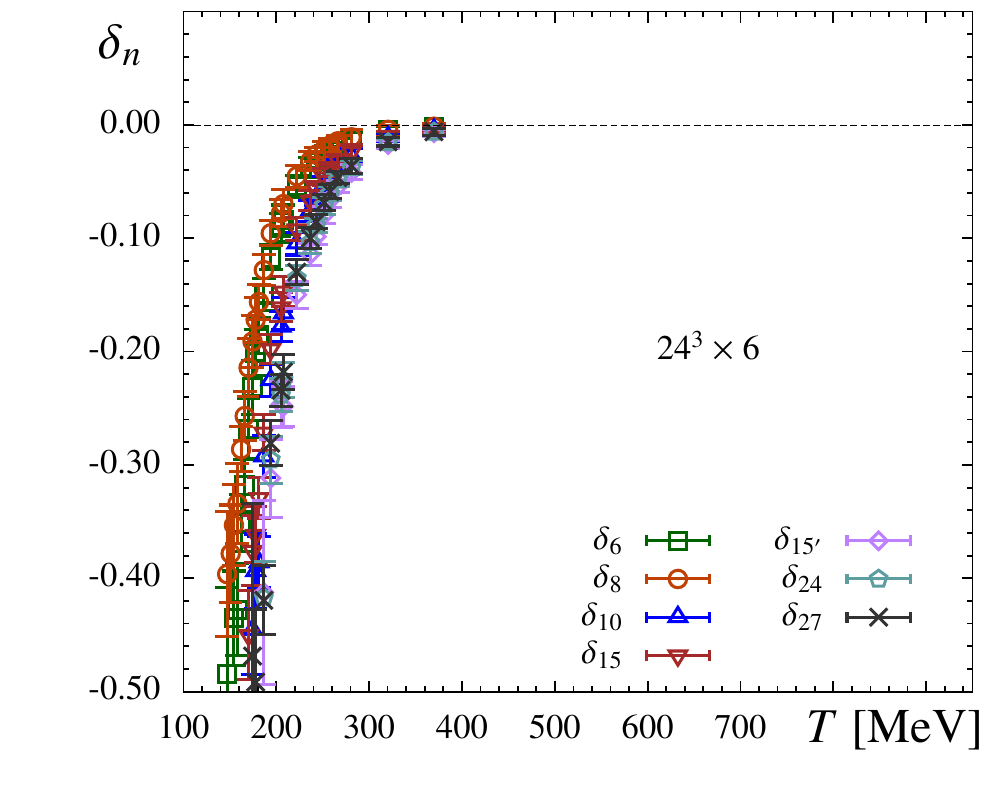}          
\includegraphics[width=5.5cm]{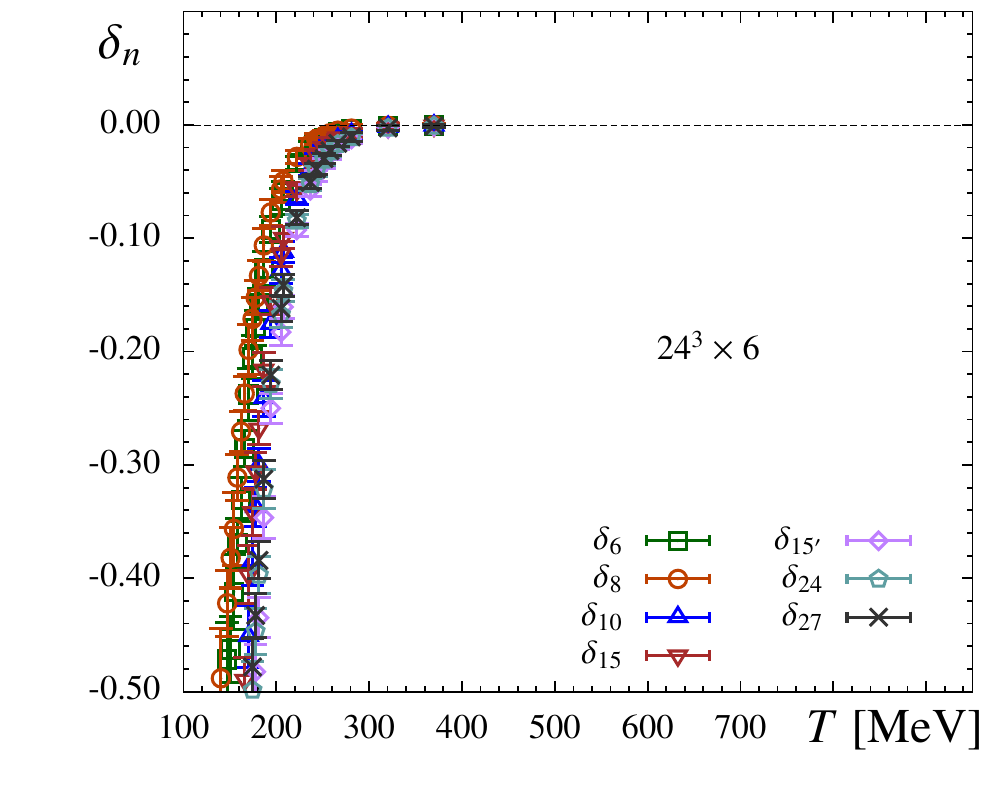}
\includegraphics[width=5.5cm]{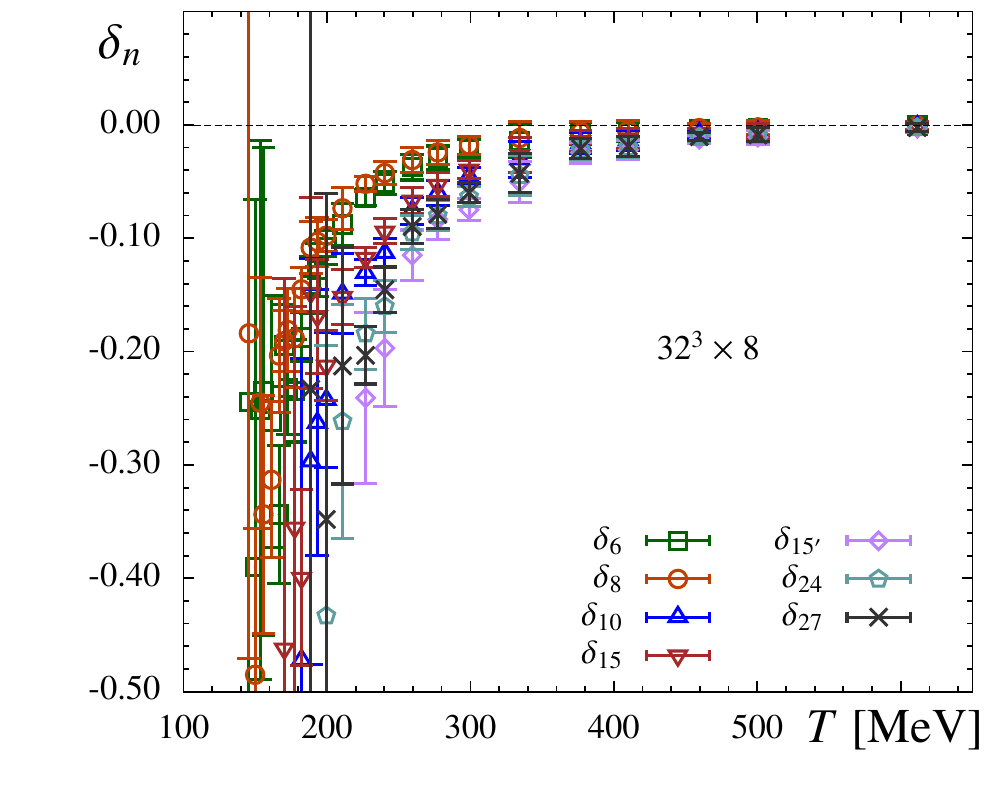}
\includegraphics[width=5.5cm]{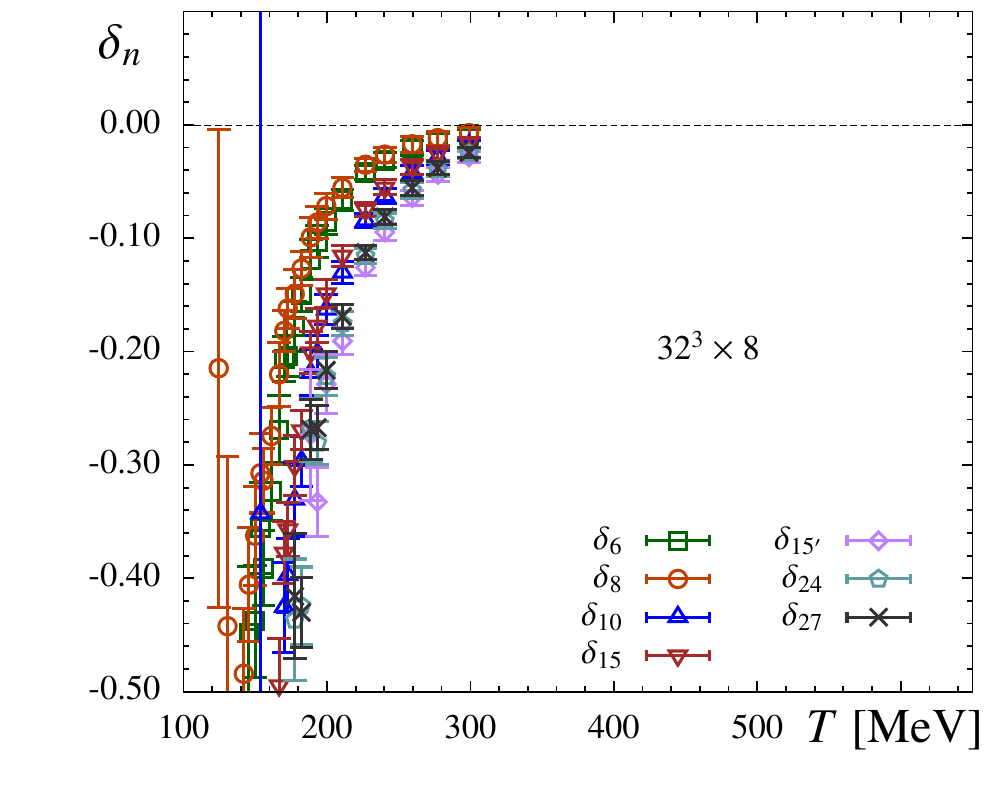}
\includegraphics[width=5.5cm]{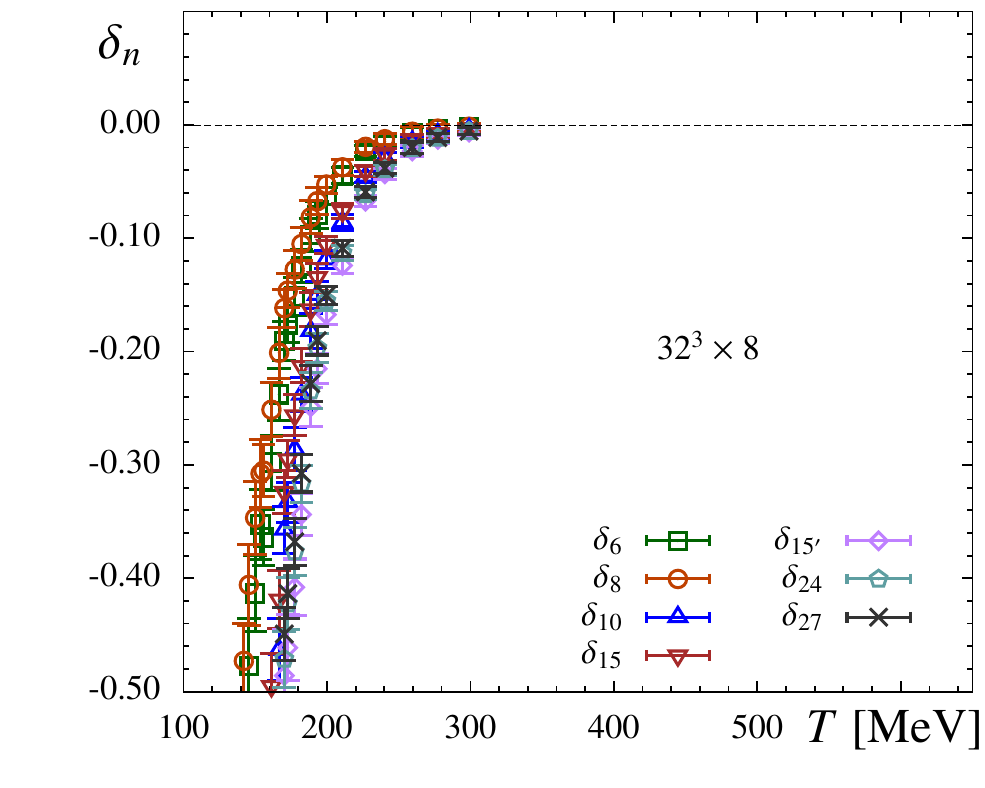}
\includegraphics[width=5.5cm]{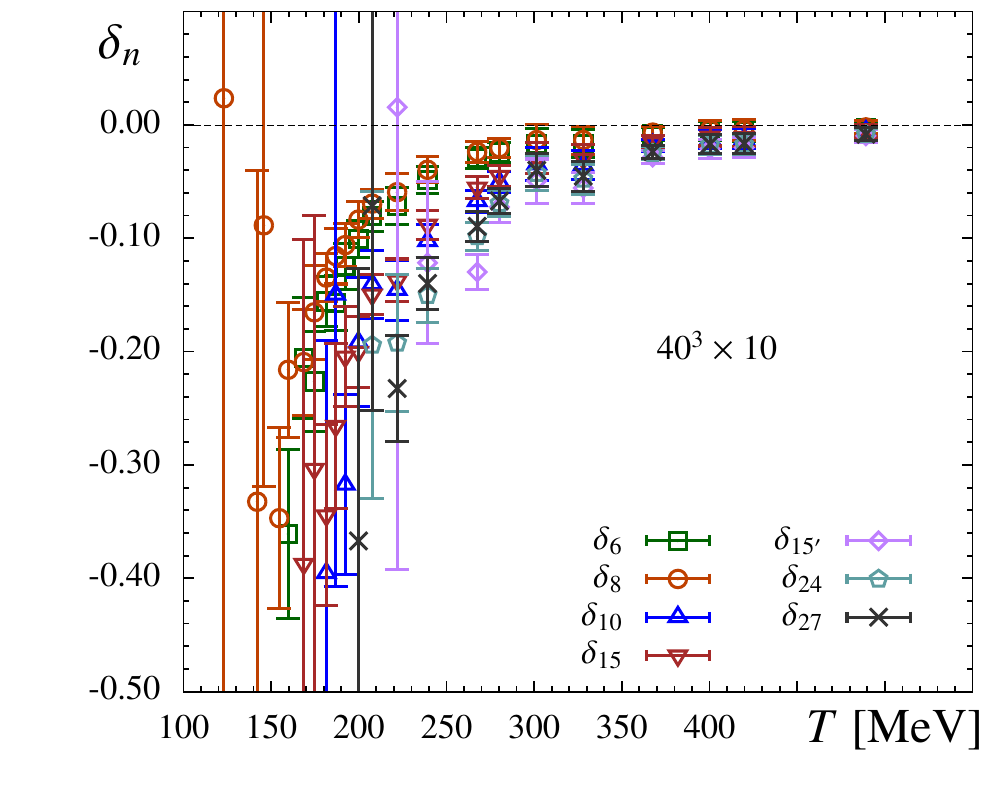}
\includegraphics[width=5.5cm]{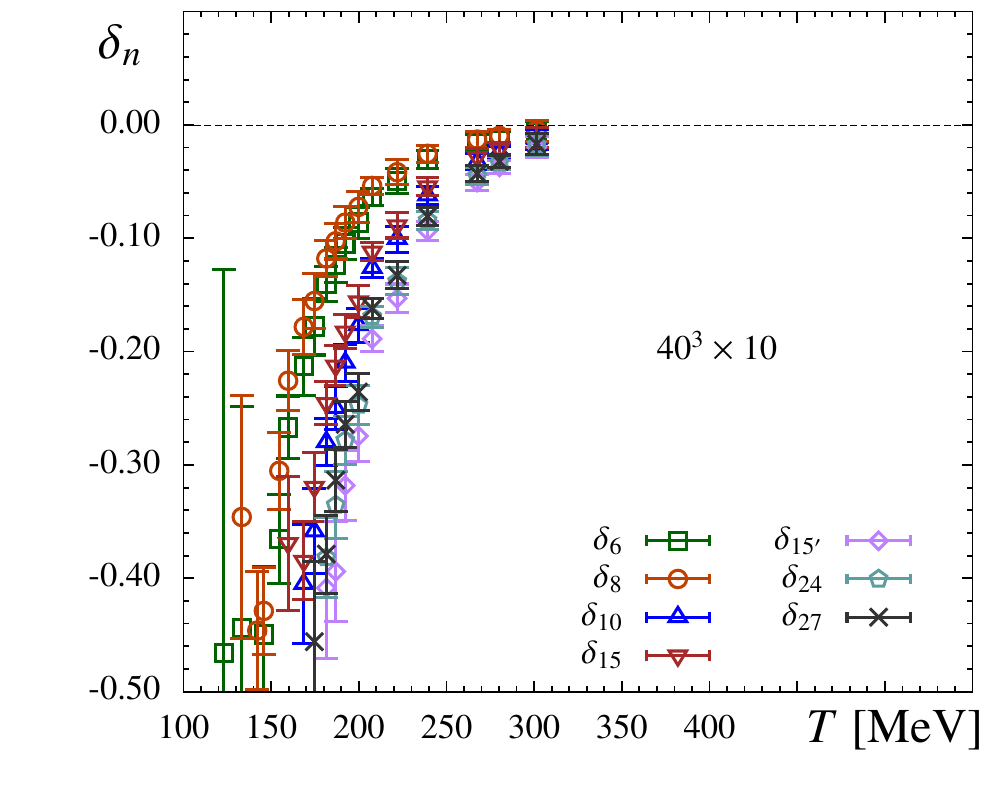}
\includegraphics[width=5.5cm]{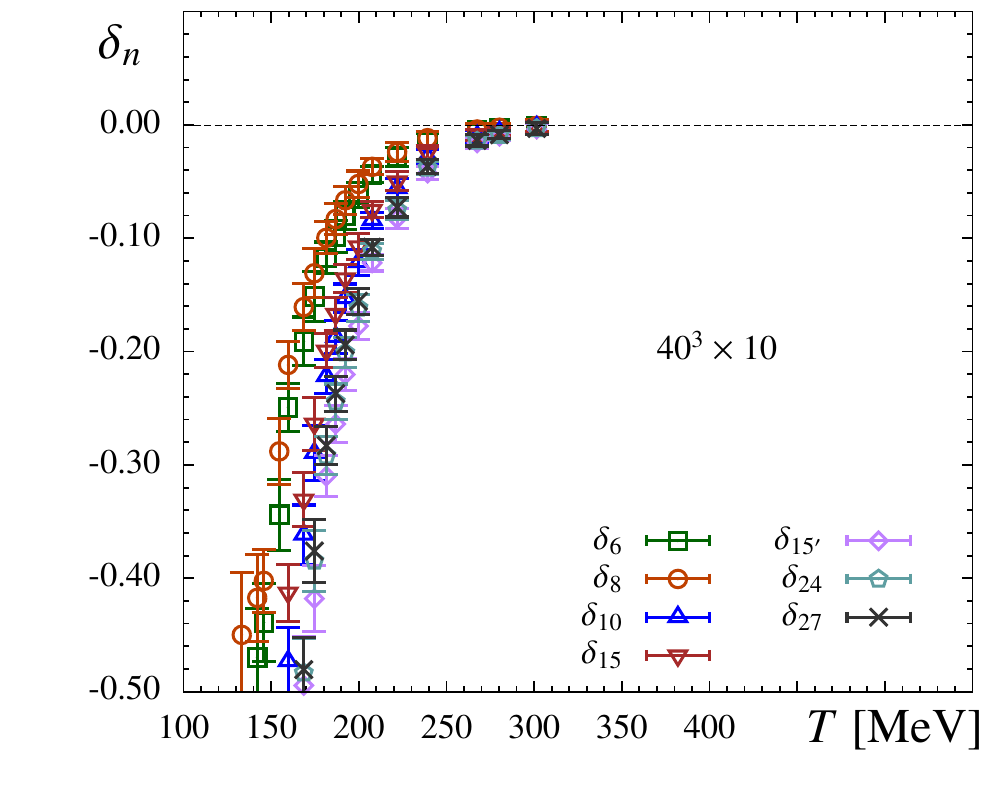}
\includegraphics[width=5.5cm]{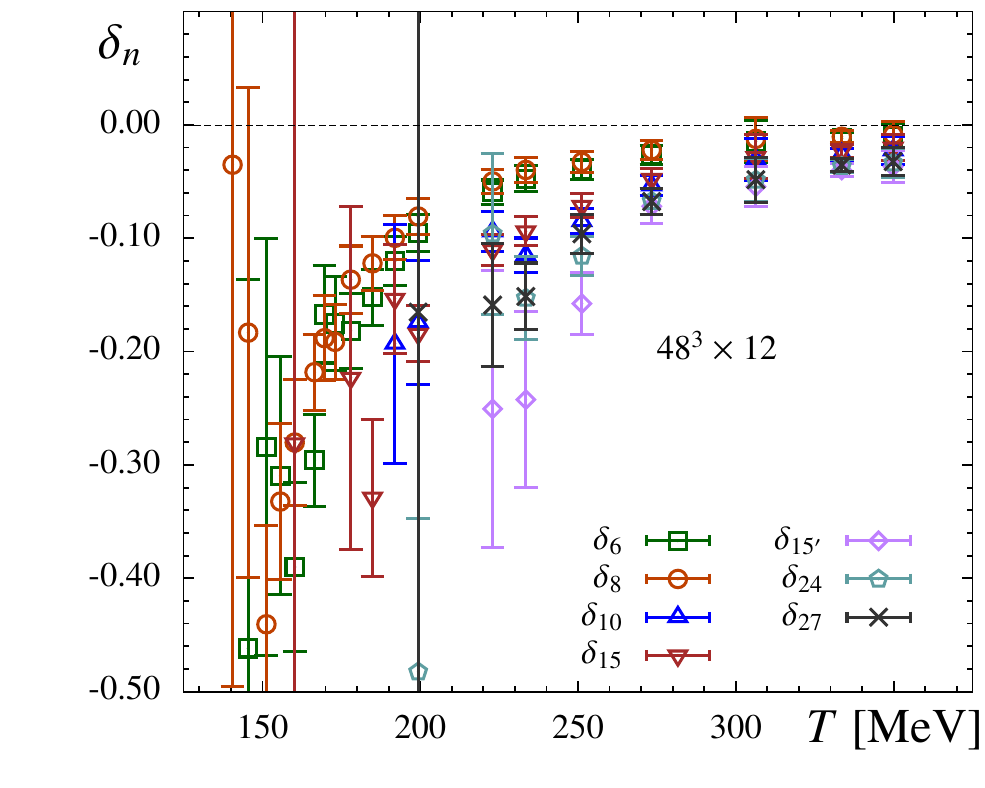}
\includegraphics[width=5.5cm]{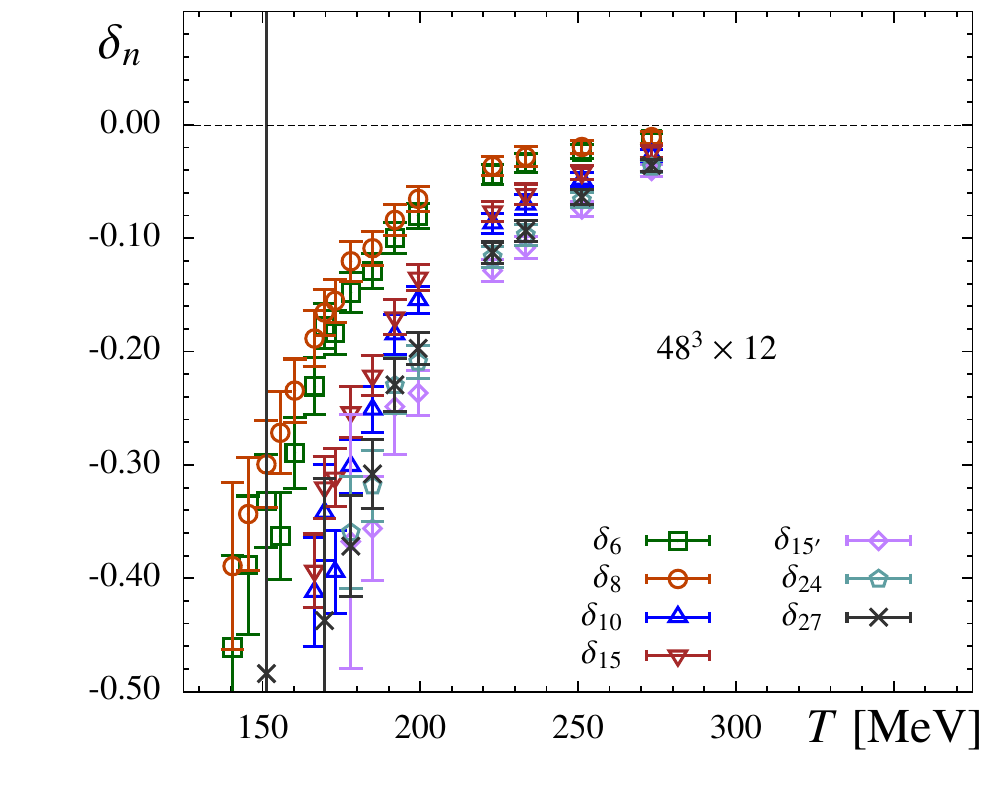}
\includegraphics[width=5.5cm]{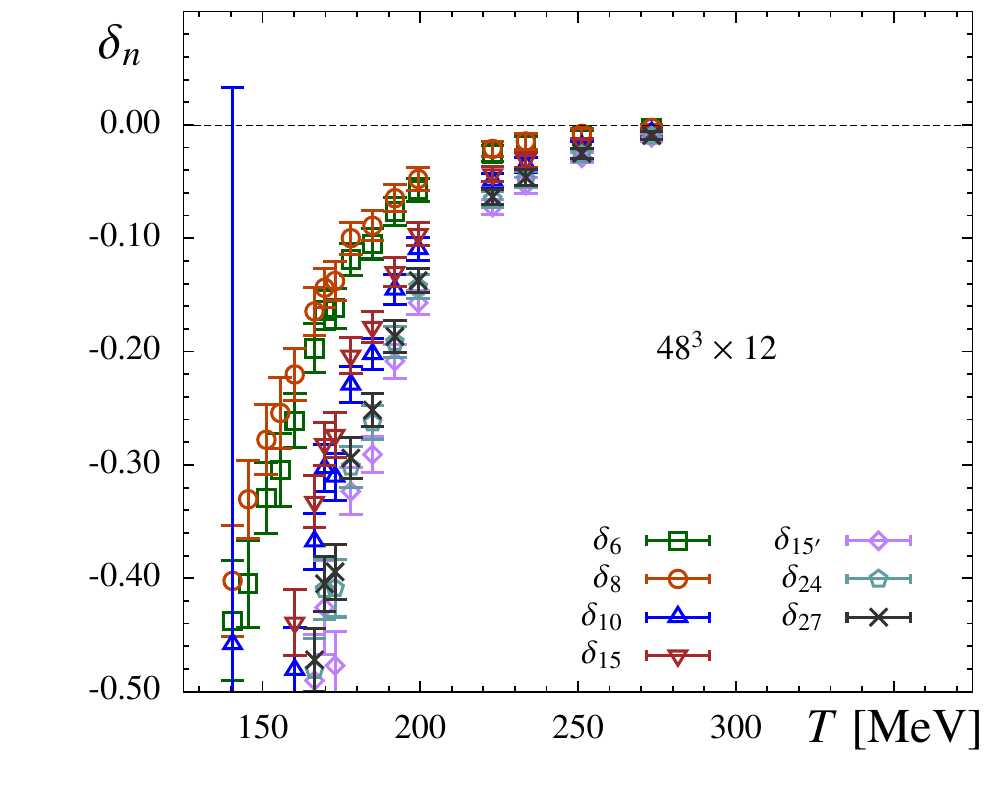}
\caption{The measure of the Casimir scaling $\delta_n$ shown for
$N_\tau=6,~8,~10$ and $12$ (from top to bottom) and flow times
$f=f_0,~2 f_0$ and $3 f_0$ (from left to right).} 
\label{fig:delta_f}
\end{figure*}

In this appendix we discuss the Casimir scaling in terms of $\delta_n$
at different flow times. In Fig. \ref{fig:delta_f} we show $\delta_n$
in various representations for $N_\tau=6,~8,~10$ and $12$. We use the values
of flow time $f=f_0,~ 2 f_0$ and $3 f_0$. For smaller values of flow time 
the data are too noisy to allow conclusive statements. From the figures
we see that the flow time dependence of $\delta_n$ is very small for
all $N_\tau$.
Therefore, we conclude that the Casimir scaling or its breaking is
independent of the flow time.

\bibliography{HotQCD}

\end{document}